\documentclass[%
 reprint,
 amsmath,amssymb,
 aps,
floatfix
]{revtex4-2}

\usepackage{graphicx}% Include figure files
\usepackage{dcolumn}% Align table columns on decimal point
\usepackage{bm}% bold math

\usepackage{hyperref}% add hypertext capabilities
\usepackage{xcolor}
\hypersetup{
    colorlinks=true,
    allcolors=blue
}
\usepackage{physics}

\usepackage{CJKutf8}


\begin{document}

\preprint{APS/123-QED}

% \title{Generalized equilibrium for color-gradient lattice Boltzmann based on higher-order Hermite polynomials: Simplified implementation in the central moments space}
\title{Generalized equilibria for color-gradient lattice Boltzmann model based on higher-order Hermite polynomials: A simplified implementation with central moments}

\author{Shimpei Saito~(\begin{CJK}{UTF8}{ipxm}齋藤慎平\end{CJK})}
 \email{s.saito@aist.go.jp}
\author{Naoki Takada~(\begin{CJK}{UTF8}{ipxm}高田尚樹\end{CJK})}
\author{\\Soumei Baba~(\begin{CJK}{UTF8}{ipxm}馬場宗明\end{CJK})}
\author{Satoshi Someya~(\begin{CJK}{UTF8}{ipxm}染矢聡\end{CJK})}
\author{Hiroshi Ito~(\begin{CJK}{UTF8}{ipxm}伊藤博\end{CJK})}
 
\affiliation{%
 Research Institute for Energy Conservation (iECO), National Institute of Advanced Industrial Science and Technology (AIST), 1-2-1 Namiki, Tsukuba 3058564, Japan
}%

\date{\today}% It is always \today, today,
             %  but any date may be explicitly specified

\begin{abstract}
We propose generalized equilibria of a three-dimensional color-gradient lattice Boltzmann model for two-component two-phase flows using higher-order Hermite polynomials.
Although the resulting equilibrium distribution function, which includes a sixth-order term on the velocity, is computationally cumbersome, its equilibrium central moments (CMs) are velocity-independent and have a simplified form.
Numerical experiments show that our approach, as in Wen \textit{et al.} [\href{https://doi.org/10.1103/PhysRevE.100.023301}{Phys. Rev. E \textbf{100}, 023301 (2019)}] who consider terms up to third order, improves the Galilean invariance compared to that of the conventional approach.
Dynamic problems can be solved with high accuracy at a density ratio of 10; however, the accuracy is still limited to a density ratio of $1\,000$.
For lower density ratios, the generalized equilibria benefit from the CM-based multiple-relaxation-time model, especially at very high Reynolds numbers, significantly improving the numerical stability.
\end{abstract}

% \keywords{Suggested keywords}%Use showkeys class option if keyword

\maketitle

% \tableofcontents

\section{Introduction}
Multiphase and multicomponent flows are ubiquitous phenomena observed in both industry and nature.
The complexity and diversity of these flows render them a fascinating subject for numerical modeling, simulation, and theoretical research.
The lattice Boltzmann (LB) method, initially proposed by~\citet{McNamara1988-nu} in 1988, has attracted attention as a powerful computational fluid dynamics tool for capturing multicomponent and multiphase flows~\citep{Aidun2010-id,Huang2015-fg,Kruger2017-ux}.
The multiphase LB model can be classified into the following categories based on the physical content of the fluid--fluid interface algorithm.
\begin{itemize}
    \item Color-gradient (CG) model~\citep{Gunstensen1991-wv,Grunau1993-ia}
    \item Pseudo-potential model~\citep{Shan1993-cx,Shan1994-wm}
    \item Free-energy model~\citep{Swift1995-wb,Swift1996-zh}
    \item Mean-field model~\citep{He1999-tm}
\end{itemize}
This classification may not be exhaustive; for instance, the free-energy and mean-field models are sometimes identified as phase-field models~\citep{Li2016-ib}.
%, since the Cahn--Hilliard or similar interface tracking equations can be derived from them.
For more details on multiphase LB models, interested readers can refer to comprehensive review papers~\citep{Chen1998-mm,Aidun2010-id,Chen2014-cs,Liu2016-ch,Li2016-ib,Hosseini2023-pf}, books~\citep{Succi2001-vt,Huang2015-fg,Kruger2017-ux,Succi2018-dc,Inamuro2021-ps} and the references therein.

Among the abovementioned models, the CG model, which introduces virtually colored distribution functions, offers many advantages in simulating multiphase and multicomponent flows.
First, the model features strict mass-conservation properties for each phase and high flexibility in setting the interfacial tension~\citep{Ba2016-ve}. 
In this model, the static drop test is not required to determine the interfacial tension, which can be obtained directly without any analysis or assumptions. 
Second, the CG model is known to have superior dissolution properties when simulating small droplets or bubbles compared to other multiphase LB models~\citep{Liu2016-ch}; i.e., the small droplets or bubbles are less prone to disappear.
In addition, the surface tension, density ratio, and viscosity ratio can be selected independently~\citep{Reis2007-xs,Kwon2023-rh}.
Studies have been conducted comparing different multiphase LB models~\citep{Huang2011-zv,Leclaire2017-rd-IntJModernPhysics,Datadien2022-tg}.
\citet{Datadien2022-tg} quantitatively compared the CG and pseudopotential models in terms of accuracy and stability and found that the CG model exhibited better characteristics.
Because of these outstanding features, the CG model has been applied to various problems related to multiphase flow, such as pseudo-boiling~\citep{Kono2000-mk}, water transport in membranes~\citep{Sarkezi-Selsky2022-tm,Sarkezi-Selsky2023-hf}, surfactant transport~\citep{liu_ba_wu_li_xi_zhang_2018,Farhat2011-ll}, thermocapillary flows~\citep{Fu2023-tj}, and droplets on microstructure~\citep{Cheng2018-tg}, etc.
Because of the compatibility of the LB method with parallel computing, the CG model has also been used for GPU-accelerated computation~\citep{Montessori2023-kh} and to acquire teacher data for machine learning~\citep{Kwon2023-rh}.
In this study, we focus on the CG model.

The CG model originates from the two-component lattice-gas automata (LGA) of \citet{Rothman1988-vm}.
\citet{Gunstensen1991-wv} proposed the first CG model by combining the single-phase LB method of \citet{McNamara1988-nu} with the two-component LGA developed by Rothman and Keller.
A perturbation step is introduced to recover Laplace's law at the interface by adding a binary fluid collision operator to the post-collision state at sites near the interface.
This model is formulated in two dimensions, and a three-dimensional model was published shortly thereafter~\citep{Gunstensen1992-zj}.
Subsequently, \citet{Grunau1993-ia} introduced different densities and viscosities by incorporating the freedom of the rest particle equilibrium distribution.
This model has been successfully applied to the two-dimensional Rayleigh--Taylor instability~\citep{Nie1998-qo}, which is known as dynamic interfacial instability.
\citet{Reis2007-xs} extended the CG model to a two-dimensional nine-velocity (D2Q9) lattice. 
They modified the perturbation operator to correctly recover the Navier--Stokes equations and showed that it could simulate the dynamic process of merging two droplets at a density ratio of 18.5.
\citet{Liu2012-fa} derived a generalized perturbation operator using the phase field (or order parameter) instead of a color gradient and described the CG model in three dimensions. 
The applicability of the CG model to dynamic problems, such as a droplet in a shear flow and a single bubble rising in a viscous fluid, was demonstrated.
\citet{Leclaire2017-kg} generalized the CG model in two and three dimensions to include solid wall boundary conditions and implemented it using the open-source PALABOS library~\citep{Latt2021-ve}.
\citet{Mora2021-ib} highlighted the importance of isotropy in the calculation of color gradients to accurately capture behaviors such as pore-scale phenomena.
Recently, \citet{Subhedar2022-lr} employed a velocity-based equilibrium function, initially proposed for the phase-field LB model~\citep{Zu2013-od}, to handle high-density ratios within the CG LB framework.

A characteristic feature of the CG model is the recoloring operation, which plays an important role in maintaining immiscibility at the interface and mimics the separation mechanism.
The original Gunstensen algorithm~\citep{Gunstensen1991-wv} is implemented by solving a maximization problem for the work performed by the color gradient against the color flux.
To reduce velocity fluctuations, \citet{Tolke2002-zs} proposed a modified algorithm in which the phase separation was not as strong as that of the Gunstensen algorithm, but the numerical stability was enhanced.
Instead of widening the interface width, Latva-Kokko—Rothman's recoloring algorithm~\citep{Latva-Kokko2005-tq}, designed based on the work of \citet{DOrtona1995-ft}, addresses certain issues with the previous CG model, namely the lattice-pinning problem and spurious currents near the fluid--fluid interface.
Subsequently, \citet{Halliday2007-mt} improved this algorithm slightly.
\citet{Subhedar2020-su} showed that Halliday's algorithm has smaller spurious currents than that of \citet{Latva-Kokko2005-tq}.
\citet{Leclaire2012-ci} integrated the Latva-Kokko--Rothmann algorithm into the Reis--Phillips CG model and demonstrated that integrating Latva-Kokko--Rothman’s recoloring operator into Reis--Phillips’ perturbation operator greatly improves the numerical stability and accuracy of solutions over a wide range of parameters. 
Using a higher-order isotropic gradient operator also enhances numerical stability and accuracy~\citep{Leclaire2011-ky}. 
Recently, the mathematical similarity between the recovered macroscopic equation from the recoloring algorithm and the conservative Allen--Cahn equation~\citep{Chiu2011-kg}, which is an interface-capturing equation used in phase-field modeling, has been discussed in several studies~\citep{Subhedar2020-su,Subhedar2022-lr,Lafarge2021-ce}.

The original CG model suffers from a lack of Galilean invariance because of the non-Navier--Stokes terms identified by \citet{Liu2012-fa}.
To restore Galilean invariance, correction terms should be added to the equilibrium distribution function~\citep{Huang2013-pd,Leclaire2013-nx}. 
Note that a similar problem exists in the free-energy model~\citep{Inamuro2000-ol,Kalarakis2002-vm,Li2021-ar}.
Following the analysis of \citet{Holdych1998-qx}, a source term for improving Galilean invariance was derived by~\citet{Leclaire2013-nx} and incorporated into an equilibrium distribution function. 
This enhanced equilibrium distribution function improved the momentum discontinuity problem through numerical tests on a layered Couette flow and was then successfully adopted in many studies~\citep{Saito2017-lg,Saito2018-ub,De_Rosis2019-bi}.
Using a different approach, \citet{Ba2016-ve} modified an equilibrium distribution function based on the third-order Hermite expansion of the Maxwellian distribution, taking a cue from \citet{Li2012-vr}. 
They also demonstrated that this modification improved the discontinuous velocity.
The CG model in Ref.~\citep{Ba2016-ve} was developed in two dimensions, whereas \citet{Wen2019-jc} later extended it to three dimensions using a D3Q19 lattice.

The collision operation in the LB method plays a pivotal role in ensuring the numerical stability and accuracy of computations~\citep{Luo2011-bw}, irrespective of whether the flow is single phase or multiphase. 
\citet{Coreixas2019-ce} have conducted an exhaustive review of collision models, presenting a systematic organization and defining the mathematical relationships among them.
From the viewpoint of computational cost, all moment spaces should be approximately equivalent since the LB scheme is memory-bound. 
This has been shown notably by \citet{Bauer2021-vl} on CPUs and by \citet{Latt2021-kd} on GPUs.
The lattice Bhatnagar--Gross--Krook (BGK) equation~\citep{Qian1992-wy,Chen1992-hd} stands out as the simplest and most widely used collision operator in the LB method. 
Based on a single relaxation time approximation, the BGK equation~\citep{Bhatnagar1954-yd} within the LB framework can be perceived as the propensity of the distribution function to gravitate toward its equilibrium state after a specified relaxation time~\citep{Kruger2017-ux}.
However, despite the success of the lattice BGK equation, it tends to become numerically unstable under the flow conditions of high Reynolds number (low-viscosity)~\citep{Ricot2009-kl}.
Several collision models have been proposed to counteract these shortcomings~\citep{Luo2011-bw,Coreixas2019-ce}.
Specifically, \citet{DHumieres1992-sp} aimed to enhance the numerical stability of the LB method by introducing a collision step in the moment space coupled with an increase in the free parameters.
This technique, which assigns different relaxation times to distinct moments, is called the multi-relaxation-time (MRT) collision model.
The MRT model has garnered considerable attention owing to its enhanced numerical stability, particularly after the publication of its formulations in two-dimensional (D2Q9)~\citep{Lallemand2000-gs} and three-dimensional (D3Q15 and D3Q19) spaces~\citep{DHumieres2002-hc}. 
In particular, determining higher-order relaxation parameters is notably challenging for three-dimensional configurations.
As a solution, the two-relaxation-time~\citep{Ginzburg2008-sh} and entropic MRT models ~\citep{Hosseini2023-sh,Hosseini2023-nj} have been proposed to offer closures for these elusive relaxation parameters.

Although the MRT model offers improved numerical stability compared to the lattice BGK model, it encounters instability at elevated Reynolds numbers. 
To address this issue and further bolster numerical stability, \citet{Geier2006-jt} introduced the cascaded LB method. This method executes the collision step in the moment space corresponding to the comoving reference frame.
Notably, in contrast to that in the raw-moment-based MRT (hereafter referred to as RM-MRT) model mentioned earlier~\citep{DHumieres1992-sp,Lallemand2000-gs,DHumieres2002-hc}, collision operations are performed in the space of the \textit{central moments} (CMs), and this is referred to as CM-MRT.
% The CM-MRT model exhibits excellent numerical stability in simulations corresponding to high Reynolds numbers~\citep{Geier2006-jt,De_Rosis2017-qm,Shan2019-wy}.
Results of several numerical simulations using the CM-MRT model have shown that it has excellent numerical stability in simulations corresponding to high Reynolds numbers~\citep{Geier2006-jt,De_Rosis2017-qm,Shan2019-wy}.
Indeed, a critical point of CM-MRT is the relaxation of (1) high-order moments and (2) the moment related to bulk viscosity. 
By imposing a relaxation frequency of 1, the stability is increased through a higher hyperviscosity and a higher bulk viscosity~\citep{Coreixas2020-pb, Wissocq2022-hy}. 
It is noted that using the same relaxation time for all moments, whether they be RM- or CM-based, leads to the same stability domain~\citep{Coreixas2020-pb}. 
The relationship between the RM- and CM-MRT models can be described under the general MRT framework in Ref.~\citep{Fei2017-rt,Fei2018-ak}.
% Recently, \citet{Luo2021-bm} introduced a unified LB model that combines multiple collision operators, facilitates seamless transitions between them and the forcing schemes, and enables practical multiphase flow simulations~\citep{Wang2022-kc,Wang2023-wz}.
The recently proposed work of \citet{Luo2021-bm} aims at a unified framework, which seamlessly integrates the widely used existing collision models. However, it would only be a first step toward a unified framework for comparing different collision models. 
For example, the matrix approach used in their work cannot reproduce the cumulant collision model~\citep{Geier2015-pv} since it is not possible to express the cumulant collision model in a linear matrix form, as pointed out in Ref.~\citep{Coreixas2019-ce}.

Because of its excellent numerical stability, CM-MRT has been successfully applied to multiphase flows~\citep{Lycett-Brown2014-ir,Leclaire2014-sw,Saito2018-ub,De_Rosis2020-uj,Cheng2021-sn,Hajabdollahi2021-pv,Saito2021-fy}.
\citet{Lycett-Brown2014-ir} pioneered the pseudo-potential model-based formulation.
In the context of the CG LB model, \citet{Leclaire2014-sw} first introduced CM-MRT in a formulation with a unit density ratio.
Subsequently, \citet{Saito2018-ub} developed a CG model with a density ratio using the nonorthogonal CM set proposed by \citet{De_Rosis2017-qm} and showed that the simulation of dynamic liquid jet flows with extremely high Reynolds numbers (up to $O(10^6)$) is possible.
To improve the potential Galilean invariance in the CG model, they used the equilibrium distribution function described by \citet{Leclaire2013-nx}.
Nevertheless, as highlighted by \citet{Wen2019-jc}, this equilibrium distribution function leaves error terms in the recovered macroscopic momentum equation.
Additionally, the functional form of the equilibrium CMs that transfers the equilibrium distribution function to the CM space is complex and cumbersome to implement.

In this study, we propose the generalization of an equilibrium distribution function to improve the Galilean invariance in the CG model using higher-order Hermite polynomials.
Furthermore, we demonstrate that the equilibrium CMs of the proposed generalized equilibrium distribution function are extremely concise and easily executable.
The remainder of this paper is organized as follows:
In Sec.~\ref{sec:color-gradient_model}, we outline the underlying D3Q27 CG LB model and describe its implementation within the framework of CM-MRT.
In Sec.~\ref{sec:equilibria}, the equilibrium distribution function within the CG LB framework is expressed in its general form using Hermite polynomials, including a comparison with existing expressions.
Furthermore, their characteristics and functional forms in the CM space are investigated.
In Sec.~\ref{sec:numerical_experiments}, through several numerical experiments, the numerical properties of the generalized equilibrium distribution function derived in this paper are presented and compared with existing distribution functions.
Sec.~\ref{sec:conclusion} concludes this paper.

\section{CG LB model\label{sec:color-gradient_model}}

\subsection{Model description}
In the present LB model, the distribution functions $f_i$ move on a D3Q27 lattice ($i \in [0,\cdots,26]$), which is a straightforward extension of the D2Q9 model~\citep{He1997-ur}, with the lattice velocity $\mathbf{c}_i$ defined as:
\begin{widetext}
\begin{equation}
\mathbf{c}_i = 
\begin{bmatrix}
    c_{ix} \\
    c_{iy} \\
    c_{iz}
\end{bmatrix}
= c
\begin{bmatrix}
   0 & 1 & -1 & 0 & 0 & 0 & 0 & 1 & -1 & 1 & -1 & 0 & 0 & 0 & 0 & 1 & -1 & 1 & -1 & 1 & -1 & 1 & -1 & 1 & -1 & -1 & 1 \\
   0 & 0 & 0 & 1 & -1 & 0 & 0 & 1 & -1 & -1 & 1 & 1 & -1 & 1 & -1 & 0 & 0 & 0 & 0 & 1 & -1 & 1 & -1 & -1 & 1 & 1 & -1 \\
   0 & 0 & 0 & 0 & 0 & 1 & -1 & 0 & 0 & 0 & 0 & 1 & -1 & -1 & 1 & 1 & -1 & -1 & 1 & 1 & -1 & -1 & 1 & 1 & -1 & 1 & -1
\end{bmatrix},
\end{equation}
\end{widetext}
where $c =\delta_x/\delta_t$, $\delta_x$ is the lattice spacing, and $\delta_t$ is the time step.
Hereafter, the formulation is given by $\delta_x = \delta_t = 1$, which is similar to the typical LB method.

In the CG LB model for two-phase flows, the two immiscible fluids are represented by introducing virtual red and blue fluids.
Distribution functions $f_i^k$ represent the fluids $k$, where $k=r$ and $b$ denote ``red'' and ``blue,'' respectively. 
The total distribution function is expressed as follows:
\begin{equation}
    f_i = f_i^r + f_i^b.
\end{equation}
The time evolution equation of the distribution function is expressed as the following LB equation with a forcing term~\citep{Guo2002-qj}:
\begin{equation}
    f_i(\vb{x}+\vb{c}_i,t+1) = f_i(\vb{x},t) + \Omega_i + F_i,
    \label{eq:LBE}
\end{equation}
where $\vb{x}$ and $t$ denote the position and time, respectively.
The last term $F_i$ introduces the body force into the LB equation; several implementations of the forcing term in the phase space have been proposed, which are reviewed in Ref.~\citep{Bawazeer2021-ig}.
The LB equation~(\ref{eq:LBE}) can be split into the collision step, described as:
\begin{equation}
    f_i^* = f_i(\vb{x},t) + \Omega_i + F_i,
    \label{eq:LBE_collision_step}
\end{equation}
and the streaming step, described as:
\begin{equation}
    f_i(\vb{x}+\vb{c}_i,t+1) = f_i^*,
    \label{eq:LBE_streaming_step}
\end{equation}
where $f_i^*$ denotes the post-collision distribution functions.
In the lattice BGK model~\citep{Qian1992-wy,Chen1992-hd}, the collision term is obtained as:
\begin{equation}
  \Omega_i = - \frac{1}{\tau} ( f_i - f_i^{\mathrm{eq}} ),
  \label{eq:collision_bgk}
\end{equation}
where $\tau$ is the relaxation time and $f_i^{\mathrm{eq}}$ is the equilibrium distribution function, which is discussed later in this paper.
The following equilibrium distribution functions are often used in standard CG LB models~\citep{Reis2007-xs, Leclaire2012-ci, Liu2012-fa, Ba2016-ve,Burgin2019-cb, Zong2021-ne, Mora2021-ib}:
\begin{equation}
  f_i^{\mathrm{eq}} = \rho \qty( \varphi_i + w_i 
    \qty[\frac{\mathbf{c}_i \cdot \mathbf{u}}{c_s^2}
    + \frac{(\mathbf{c}_i \cdot \mathbf{u})^2}{2c_s^4}
    - \frac{|\mathbf{u}|^2}{2c_s^2}]
    ), \label{eq:standard_CG_2nd}
\end{equation}
where $c_s$ is the speed of sound defined by the LB method for standard single-phase flows ($c_s^2=1/3$ for the D3Q27 lattice~\citep{He1997-ur}) and $\varphi_i$ is the lattice-specific parameter, which can be expressed as:
% \begin{equation}
%     \varphi_i = 1 - \frac{p}{\rho} \qty( \frac{1}{\xi} \qty(1-|\vb{c}_i|^2) )
% \end{equation}
\begin{equation}
    \varphi_i = 
    \begin{cases}
        ~1 - \dfrac{1}{\xi} \dfrac{p}{\rho}, & |\vb{c}_i| = 0\\
        ~w_i \dfrac{p}{\rho c_s^2}. & |\vb{c}_i| \neq 0
    \end{cases}
\end{equation}
For the D3Q27 lattice, $\xi = 9/19$, and $\xi$ values for other lattice models can be found in Ref.~\citep{Leclaire2017-kg}.
The weight function is given by: 
\begin{equation}
    w_i = 
    \begin{cases}
        ~8/27, & |\vb{c}_i| = 0 \\
        ~2/27, & |\vb{c}_i| = 1 \\
        ~1/54, & |\vb{c}_i| = \sqrt{2} \\
        ~1/216. & |\vb{c}_i| = \sqrt{3} \\
    \end{cases}
\end{equation}
The density of the $k$-phase fluid is given by:
\begin{equation}
    \rho_k = \sum_i f_i^k.
\end{equation}
The total fluid density is given by $\rho = \sum_k \rho_k$.
The momentum is defined as follows: 
\begin{equation}
    \rho \mathbf{u} = \sum_i f_i \mathbf{c}_i + \frac{\mathbf{F}}{2}, 
\end{equation}
where $\mathbf{F}$ denotes the body force.

Excluding a few recent CG models~\citep{Lafarge2021-ce,Subhedar2022-lr}, the pressure $p$ and density $\rho$ are connected by an ideal gas equation of state (EOS). 
To obtain a continuous pressure at the interface, the CG model introduces different sound speeds for each phase.
Using the $k$-phase speed of sound $c_s^k$, the pressure can be expressed as:
\begin{equation}
    p = \sum_k p_k = \sum_k{\rho_k} (c_s^k)^2,
    \label{eq:pressure}
\end{equation}
where $p_k = \rho_k (c_s^k)^2$ is the pressure of $k$-phase.
This feature results in different sound speeds being defined for both phases.
From the pressure balance $p_r=p_b$ in the interface region, we obtain the following relationship between the density ratio and speed of sound~\citep{Reis2007-xs, Burgin2019-cb,Lafarge2021-ce,Spendlove2020-mx}:
\begin{equation}
    \gamma = 
    \frac{\rho_r^0}{\rho_b^0} 
    = \qty(\frac{c_s^b}{c_s^r})^2
    = \frac{1-\alpha_b}{1-\alpha_r},
    \label{eq:density_ratio_vs_speed_of_sounds}
\end{equation}
where the superscript ``0'' indicates the initial density value of the simulation~\citep{Leclaire2013-nx} and the relation $(c_s^k)^2 = \xi \rho_k(1-\alpha_k)$ holds.
Eq.~(\ref{eq:density_ratio_vs_speed_of_sounds}) implies that the greater the density ratio, the greater the difference in the speed of sound between the two phases; the speed of sound in the liquid phase becomes progressively smaller than that in the gas phase.
Note that to overcome such limitations, \citet{Lafarge2021-ce} reformulate the pressure definition under the assumption of mechanical equilibrium~\citep{Saurel2016-es,Boivin2019-ba} and successfully introduced a mixture EOS for two-phase flow, inspired by the stiffened gas formulation~\citep{Le_Metayer2004-ov}, into the CG model.

The fluid interface is tracked using an order parameter that distinguishes the two components in a multicomponent flow, defined as~\citep{Ba2016-ve, Saito2018-ub}:
\begin{equation}
    \phi(\vb{x},t) = \frac{\rho_r(\vb{x},t)/\rho_r^0 - \rho_b(\vb{x},t)/\rho_b^0}{\rho_r(\vb{x},t)/\rho_r^0 + \rho_b(\vb{x},t)/\rho_b^0}.
    \label{eq:order_parameter}
\end{equation}
The order parameter values $\phi=1$, $-1$, and $0$ correspond to a purely red fluid, a purely blue fluid, and the interface between the two, respectively~\citep{Tolke2002-zs}.
The interfacial tension between the two fluids is introduced as a spatially varying body force $\mathbf{F}_s$ based on the continuum surface force (CSF)~\citep{Brackbill1992-ab}, which is defined as:
\begin{equation}
    \mathbf{F}_s = \frac{1}{2} \sigma \kappa \nabla \phi, \label{eq:CSF_force}
\end{equation}
where $\sigma$ is the interfacial tension coefficient and $\kappa$ is the interface curvature~\citep{Lishchuk2003-aw}, which is expressed as:
\begin{equation}
    \kappa = -\nabla_s \cdot \vb{n},
\end{equation}
where the surface gradient operator $\nabla_s = ( \vb{I} - \vb{n}\vb{n} )\cdot \nabla$.
Using the order parameters in Eq.~(\ref{eq:order_parameter}), the unit normal vector is defined as:%$\vb{n} = \nabla \phi / |\nabla \phi|$
\begin{equation}
    \vb{n} = \frac{\nabla \phi}{|\nabla \phi|}.
\end{equation}

A recoloring step is introduced to maintain immiscibility in the interfacial region, which is a feature of the CG LB model.
Following the algorithm developed by \citet{Halliday2007-mt}, the recoloring operation can be represented as:
% \begin{equation}
%     \begin{split}
%         f_i^r = & ~ \frac{\rho_r}{\rho} f_i 
%     + \beta w_i \frac{\rho_r \rho_b}{\rho^2}  \frac{p}{c_s^2} (\mathbf{c}_i \cdot \vb{n}), \\
%   % blue
%         f_i^b = & ~ \frac{\rho_b}{\rho} f_i 
%     - \beta w_i \frac{\rho_r \rho_b}{\rho^2} \frac{p}{c_s^2} (\mathbf{c}_i \cdot \vb{n}),
%     \end{split}
% \end{equation}
\begin{equation}
    \begin{split}
        f_i^{r,*} &= \frac{\rho_r}{\rho} f_i^* 
        + \frac{ w_i}{c_s^2} \vb{c}_i \cdot \vb{R}, \\
  % blue
        f_i^{b,*} &= \frac{\rho_b}{\rho} f_i^* 
    - \frac{ w_i}{c_s^2} \vb{c}_i \cdot \vb{R}, \label{eq:recoloring}
    \end{split}
\end{equation}
with
\begin{equation}
    \vb{R} = \beta \frac{\rho_r \rho_b}{\rho^2}p\vb{n},
    \label{eq:recoloring_force}
\end{equation}
where $\beta$ is the parameter that controls the interface thickness.
In Eq.~(\ref{eq:recoloring_force}), the pressure $p$ comes from the equilibrium at rest as usually found in the literature (e.g.~\citep{Leclaire2017-kg, Saito2018-ub}).
Unless otherwise stated, we set $\beta=0.7$ to reproduce the correct interfacial behavior with as narrow an interface as possible~\citep{Liu2012-fa,Halliday2007-mt,Liu2017-fd}.
After summing $f_i^{r,*}$ and $f_i^{b,*}$, the streaming operation is performed according to Eq.~(\ref{eq:LBE_streaming_step}) and the boundary conditions are implemented as necessary.

The partial derivatives of the variable $\chi$ are evaluated using the second-order isotropic finite difference~\citep{Liu2012-fa, Guo2011-mh, Lou2012-fi}:
\begin{equation}
    \nabla \chi(\vb{x},t) = \frac{1}{c_s^2} \sum_i w_i \chi(\vb{x}+\vb{c}_i,t)\vb{c}_i,
    \label{eq:second-order_isotropic_finite_difference}
\end{equation}
To ensure smoothness across the interface, the kinematic viscosity is interpolated using the following order parameter:
\begin{equation}
    \nu(\vb{x},t) = \frac{1}{2} \qty(1+\phi(\vb{x},t)) \nu_r + \frac{1}{2} \qty(1-\phi(\vb{x},t)) \nu_b,
\end{equation}
where $\nu_k$ denotes the $k$-phase kinematic viscosity.
From the relationship between the relaxation time and viscosity~\citep{Ba2016-ve}, we obtain:
\begin{equation}
    \mu = \qty(\tau - \frac{1}{2})p,
    \label{eq:relation_viscosity_relaxation}
\end{equation}
where $\mu = \rho \nu$ denotes dynamic viscosity.

\subsection{CM-MRT collision operation}
In this study, the collision step in Eq.~(\ref{eq:LBE_collision_step}) is implemented in the CM space.
To achieve this, it is necessary to transform the distribution functions into CMs.
This procedure first converts the distribution functions into RMs, which are further converted into CMs.
The RMs and CMs are defined as~\citep{Geier2006-jt}:
\begin{align}
    m_{\alpha\beta\gamma} &= \sum_i {
    f_i 
    c_{ix}^\alpha
    c_{iy}^\beta
    c_{iz}^\gamma}, \label{eq:definition_raw_moments} \\
    k_{\alpha\beta\gamma} &= \sum_i {
    f_i 
    (c_{ix} - u_x)^\alpha
    (c_{iy} - u_y)^\beta
    (c_{iz} - u_z)^\gamma},
    \label{eq:definition_central_moments}
\end{align}
where $\alpha$, $\beta$, and $\gamma$ range from zero to two, yielding $3^3=27$ moments.
Although the CMs can be computed directly from Eq.~(\ref{eq:definition_central_moments}), fewer computations are required when using the RMs [Eq.~(\ref{eq:definition_raw_moments})].
Specific conversion expressions between the distribution functions, RMs, and CMs can be obtained by executing \texttt{get\_f\_m\_k\_relations.jl}, as presented in the Supplementary Material.

% \textcolor{red}{According to the procedure and notation described in \citet{Geier2015-pv}, the collision step [Eq.~(\ref{eq:LBE_collision_step})] in the CM space with the forcing terms~\citep{De_Rosis2019-bi,Fei2018-ak,Luo2021-bm} can be represented as:
In this paper, the collision step in the CM space is based on Appendix D of \citet{Geier2015-pv};
however, the correction terms added to the second-order moment collisions in  Ref.~\citep{Geier2015-pv} are not considered because they are based on a Taylor expansion that falls apart close to strong discontinuities such as walls and phase interfaces.
It also differs from Ref.~\citep{Geier2015-pv} in that it considers third- and fifth-order forcing terms, in accordance with Refs.~\citep{De_Rosis2019-bi,Fei2018-ak,Luo2021-bm}.
As a result, the collision step can be expressed as:
\begin{widetext}
\begin{equation}
  \begin{split}
  % 0th and 2nd order
  % k_{000}^{\mathrm{eq}} = &~ \rho, \\
  % k_{100}^{*} = &~ (1 - \omega_0)k_{100} + \qty(1 - \frac{\omega_0}{2})k_{100}^{\mathrm{F}}  ,  \\
  % k_{010}^{*} = &~ (1 - \omega_0)k_{010} + \qty(1 - \frac{\omega_0}{2})k_{010}^{\mathrm{F}}  ,  \\
  % k_{001}^{*} = &~ (1 - \omega_0)k_{001} + \qty(1 - \frac{\omega_0}{2})k_{001}^{\mathrm{F}}  ,  \\
  k_{100}^{*} &= (1 - \omega_0)k_{100} + \qty(1-\omega_0/2)F_x ,  \\
  k_{010}^{*} &= (1 - \omega_0)k_{010} + \qty(1-\omega_0/2)F_y ,  \\
  k_{001}^{*} &= (1 - \omega_0)k_{001} + \qty(1-\omega_0/2)F_z ,  \\
  k_{110}^{*} &= (1 - \omega_1)k_{110},  \\
  k_{011}^{*} &= (1 - \omega_1)k_{011},  \\
  k_{101}^{*} &= (1 - \omega_1)k_{101}  ,  \\
  k_{200}^{*} - k_{020}^{*} &= (1 - \omega_1)(k_{200} - k_{020}) + \omega_1 (k_{200}^\mathrm{eq} - k_{020}^\mathrm{eq})
  + \qty(1-\omega_1/2)(Q_x - Q_y),  \\
  k_{200}^{*} - k_{002}^{*} &= (1 - \omega_1)(k_{200} - k_{002}) + \omega_1 (k_{200}^\mathrm{eq} - k_{002}^\mathrm{eq})
  + \qty(1-\omega_1/2)(Q_x - Q_z),  \\
  k_{200}^{*} + k_{020}^{*} + k_{002}^{*} &= (1 - \omega_2)(k_{200} + k_{020} + k_{002}) + \omega_2 (k_{200}^\mathrm{eq} + k_{020}^\mathrm{eq} + k_{002}^\mathrm{eq})
  + \qty(1-\omega_2/2)(Q_x + Q_y + Q_z),  \\
  % 3rd order
  % k_{120}^{*} + k_{102}^{*} = & ~(1-\omega_3)(k_{120} + k_{102}) 
  % + \omega_3 (k_{120}^\mathrm{eq} + k_{102}^\mathrm{eq}) 
  % + \qty(1 - \frac{\omega_3}{2})(k_{120}^{\mathrm{F}} + k_{102}^{\mathrm{F}}), \\
  % k_{210}^{*} + k_{012}^{*} = & ~(1-\omega_3)(k_{210} + k_{012}) 
  % + \omega_3 (k_{210}^\mathrm{eq} + k_{012}^\mathrm{eq})
  % + \qty(1 - \frac{\omega_3}{2})(k_{210}^{\mathrm{F}} + k_{012}^{\mathrm{F}}), \\
  % k_{201}^{*} + k_{021}^{*} = & ~(1-\omega_3)(k_{201} + k_{021}) 
  % + \omega_3 (k_{201}^\mathrm{eq} + k_{021}^\mathrm{eq})
  % + \qty(1 - \frac{\omega_3}{2})(k_{201}^{\mathrm{F}} + k_{021}^{\mathrm{F}}), \\
  % %%%%
  % k_{120}^{*} - k_{102}^{*} = & ~(1-\omega_4)(k_{120} - k_{102}) 
  % + \omega_4 (k_{120}^\mathrm{eq} - k_{102}^\mathrm{eq})
  % + \qty(1 - \frac{\omega_4}{2})(k_{120}^{\mathrm{F}} - k_{102}^{\mathrm{F}}), \\
  % k_{210}^{*} - k_{012}^{*} = & ~(1-\omega_4)(k_{210} - k_{012}) 
  % + \omega_4 (k_{210}^\mathrm{eq} - k_{012}^\mathrm{eq})
  % + \qty(1 - \frac{\omega_4}{2})(k_{210}^{\mathrm{F}} - k_{012}^{\mathrm{F}}), \\
  % k_{201}^{*} - k_{021}^{*} = & ~(1-\omega_4)(k_{201} - k_{021}) 
  % + \omega_4 (k_{201}^\mathrm{eq} - k_{021}^\mathrm{eq})
  % + \qty(1 - \frac{\omega_4}{2})(k_{201}^{\mathrm{F}} - k_{021}^{\mathrm{F}}), \\
  k_{120}^{*} + k_{102}^{*} &= (1-\omega_3)(k_{120} + k_{102}) 
  + \omega_3 (k_{120}^\mathrm{eq} + k_{102}^\mathrm{eq}) 
  + 2\qty(1-\omega_3/2)F_x c_s^2, \\
  k_{210}^{*} + k_{012}^{*} &= (1-\omega_3)(k_{210} + k_{012}) 
  + \omega_3 (k_{210}^\mathrm{eq} + k_{012}^\mathrm{eq})
  + 2\qty(1-\omega_3/2)F_y c_s^2, \\
  k_{201}^{*} + k_{021}^{*} &= (1-\omega_3)(k_{201} + k_{021}) 
  + \omega_3 (k_{201}^\mathrm{eq} + k_{021}^\mathrm{eq})
  + 2\qty(1-\omega_3/2)F_z c_s^2, \\
  %%%%
  k_{120}^{*} - k_{102}^{*} &= (1-\omega_4)(k_{120} - k_{102}) 
  + \omega_4 (k_{120}^\mathrm{eq} - k_{102}^\mathrm{eq}), \\
  k_{210}^{*} - k_{012}^{*} &= (1-\omega_4)(k_{210} - k_{012}) 
  + \omega_4 (k_{210}^\mathrm{eq} - k_{012}^\mathrm{eq}), \\
  k_{201}^{*} - k_{021}^{*} &= (1-\omega_4)(k_{201} - k_{021}) 
  + \omega_4 (k_{201}^\mathrm{eq} - k_{021}^\mathrm{eq}), \\
  k_{111}^{*}               &= (1-\omega_5) k_{111} + \omega_5 k_{111}^\mathrm{eq}, \\
  % 4th order
  k_{220}^{*} - 2k_{202}^{*} + k_{022}^{*} &=  (1-\omega_6)(k_{220} - 2k_{202} + k_{022})
  + \omega_6 (k_{220}^\mathrm{eq} - 2k_{202}^\mathrm{eq} + k_{022}^\mathrm{eq}),  \\
  k_{220}^{*} + k_{202}^{*} - 2k_{022}^{*}  &= (1-\omega_6)(k_{220} + k_{202} - 2k_{022})
  + \omega_6 (k_{220}^\mathrm{eq} + k_{202}^\mathrm{eq} - 2k_{022}^\mathrm{eq}),  \\
  k_{220}^{*} + k_{202}^{*} + k_{022}^{*} &= (1-\omega_7)(k_{220} + k_{202} + k_{022}) 
  + \omega_7 (k_{220}^\mathrm{eq} + k_{202}^\mathrm{eq} + k_{022}^\mathrm{eq}),  \\
  k_{211}^{*} &= (1-\omega_8)k_{211} + \omega_8 k_{211}^\mathrm{eq}, \\
  k_{121}^{*} &= (1-\omega_8)k_{121} + \omega_8 k_{121}^\mathrm{eq}, \\
  k_{112}^{*} &= (1-\omega_8)k_{112} + \omega_8 k_{112}^\mathrm{eq}, \\
  % Fifth order:
  % k_{122}^{*} = & ~ (1-\omega_9)k_{122} + \omega_9 k_{122}^\mathrm{eq} 
  % + \qty(1 - \frac{\omega_9}{2})k_{122}^{\mathrm{F}}, \\
  % k_{212}^{*} = & ~ (1-\omega_9)k_{212} + \omega_9 k_{212}^\mathrm{eq}
  % + \qty(1 - \frac{\omega_9}{2})k_{212}^{\mathrm{F}}, \\
  % k_{221}^{*} = & ~ (1-\omega_9)k_{221} + \omega_9 k_{221}^\mathrm{eq}
  % + \qty(1 - \frac{\omega_9}{2})k_{221}^{\mathrm{F}}, \\
  k_{122}^{*} &= (1-\omega_9)k_{122} + \omega_9 k_{122}^\mathrm{eq} 
  + \qty(1 - \omega_9/2)F_x c_s^4, \\
  k_{212}^{*} &= (1-\omega_9)k_{212} + \omega_9 k_{212}^\mathrm{eq}
  + \qty(1 - \omega_9/2)F_y c_s^4, \\
  k_{221}^{*} &= (1-\omega_9)k_{221} + \omega_9 k_{221}^\mathrm{eq}
  + \qty(1 - \omega_9/2)F_z c_s^4, \\
  % 6th order
  k_{222}^{*} &= (1-\omega_{10})k_{222}  + \omega_{10} k_{222}^\mathrm{eq}, 
  \label{eq:collision_step_in_general_form}
  \end{split}
\end{equation}
\end{widetext}
where $k_{\alpha\beta_\gamma}^*$ and $k_{\alpha\beta_\gamma}^{\mathrm{eq}}$ are the post-collision and equilibrium CMs, respectively.
The equilibrium CMs are described in Sec. ~\ref{sec:equilibria}.
$\omega_0$, $\omega_1$, $\cdots$, $\omega_{10}$ are the relaxation rates.
$\omega_1$ is related to the kinematic viscosity based on the relation $\omega_1 = 1/\tau$ and Eq.~(\ref{eq:relation_viscosity_relaxation}), and $\omega_2$ is associated with the bulk viscosity.
The others are free and can be chosen from the range $\{0 \cdots 2 \}$.
The correction terms $Q_x$, $Q_y$, and $Q_z$ appearing in the second-order collision stages are specified in Sec. ~\ref{sec:Correction_for_the_diagonal_elements}.

Following the collision operation in the CM space, the CMs are transformed back into the equilibrium distribution functions via the RMs.
The specific conversion expressions between them can be obtained by executing \texttt{get \_f\_m\_k\_relations.jl}, as described in the Supplementary Material.

\section{Equilibria for CG LB \label{sec:equilibria}}
Before discussing equilibria in the CG model, we briefly review equilibria for single-phase flows.
Within this framework, the most widely used equilibrium distribution function can be represented as~\citep{Qian1992-wy,Chen1998-mm}:
\begin{equation}
    g_i^{\mathrm{eq},2} = \rho w_i \qty( 1 +  
    \frac{\mathbf{c}_i \cdot \mathbf{u}}{c_s^2}
    + \frac{(\mathbf{c}_i \cdot \mathbf{u})^2}{2c_s^4}
    - \frac{|\mathbf{u}|^2}{2c_s^2}
    ). \label{eq:equilibria_for_standard_LBM_2nd}
\end{equation}
This equation is obtained by Taylor expanding the Maxwell--Boltzmann distribution with macroscopic velocities assuming a low Mach number and considering terms up to $O(u^2)$.
In contrast, \citet{De_Rosis2019-bq} expanded the Maxwell--Boltzmann distribution to an arbitrary order $O(u^N)$ using Hermite polynomials~\citep{Coreixas2017-my,Shan2006-tf}. 
The relaxation of the continuous Maxwellian distribution is equivalent to that of  its discrete counterpart when the equilibrium state is constructed using sixth-order Hermite polynomials in three dimensions~\citep{De_Rosis2019-bq}, described as follows:
\begin{widetext}
\begin{equation}
  \begin{split}
    g_i^{\mathrm{eq},6} = & ~ \rho w_i \left[  1 + 
    %% 1st order
      \frac{u_x H_{i100} + u_y H_{i010} + u_z H_{i001}}{c_s^2} \right.  \\
    %% 2nd order 
    & \left. 
    + \frac{u_x^2 H_{i200} + u_y^2 H_{i020} + u_z^2 H_{i002} 
      + 2( 
          u_x u_y H_{i110} 
        + u_y u_z H_{i011}
        + u_x u_z H_{i101}
      )}{2c_s^4} \right.\\
    %% 3rd order
    & \left.  + \frac{u_x^2 u_y H_{i210} 
      + u_x^2 u_z H_{i201} 
      + u_x u_y^2 H_{i120} 
      + u_x u_z^2 H_{i102} 
      + u_y u_z^2 H_{i012} 
      + u_y^2 u_z H_{i021}  
      + 2u_x u_y u_z H_{i111}}{2c_s^6} \right.\\
    %% 4th order
      & \left. 
      + \frac{u_x^2 u_y^2 H_{i220}
        + u_x^2 u_z^2 H_{i202}
        + u_y^2 u_z^2 H_{i022}
        + 2(
            u_x u_y u_z^2 H_{i112}
          + u_x u_y^2 u_z H_{i121}
          + u_x^2 u_y u_z H_{i211})}{4c_s^8} \right.\\
    %% Fifth order:
    & \left. 
      + \frac{u_x^2 u_y u_z^2 H_{i212} 
        + u_x^2 u_y^2 u_z H_{i221}  
        + u_x u_y^2 u_z^2 H_{i122}}{4c_s^{10}} %\right. \right.\\
    %% 6th order
    %& \left. \left. 
      + \frac{u_x^2 u_y^2 u_z^2  H_{i222}}{8c_s^{12}}  
      \right], \label{eq:equilibria_for_standard_LBM_6th}
  \end{split}
\end{equation}
\end{widetext}
where $H_{i\alpha\beta\gamma}$ denotes the Hermite polynomial presented in Appendix~\ref{sec:Appendix_Hermite_polynomials}.
When Hermite polynomials of orders of three and higher are neglected, Eq.~(\ref{eq:equilibria_for_standard_LBM_6th}) can be reduced to Eq.~(\ref{eq:equilibria_for_standard_LBM_2nd}).

We now return to the CG LB model.
The equilibrium distribution function within the CG framework expressed by Eq.~(\ref{eq:standard_CG_2nd}) is generally adopted when dealing with density contrasts.
In this study, we reformulate the equilibrium distribution function for the CG model as:
\begin{equation}
    f_i^{\mathrm{eq}} = g_i^{\mathrm{eq},N} + (p-\rho c_s^2) (E_i + \Phi_i), 
    \label{eq:equilibria_in_general_form}
\end{equation}
where $E_i$ is an isotropic operator that ensures interface isotropy~\citep{Lafarge2021-ce}, which can be expressed as:
\begin{equation}
  \begin{split}
      E_i = w_i 
      & 
      \left(\frac{H_{i200}+H_{i020}+H_{i002}}{2c_s^4} \right. \\
      &  \left. - \frac{H_{i220}+H_{i022}+H_{i202}}{4c_s^6}
      + \frac{H_{i222}}{8c_s^8}
      \right).
      \label{eq:isotropic_operator}
  \end{split}
  \end{equation}
The correction operator $\Phi_i$ is introduced to recover the Galilean invariance.
The specific functional form of the correction operator $\Phi_i$ is defined in this section.
The second term of the RHS in Eq.~(\ref{eq:equilibria_in_general_form}) represents the deviation from the ideal gas; that is, when $p = \rho c_s^2$ for each phase (e.g., unit density ratio), the second term can be neglected and the formulation for the usual single-phase flow can be recovered.

The functional form of equilibrium CMs is also an important aspect in the following discussion.
As depicted in Eq.~(\ref{eq:definition_central_moments}), the equilibrium CMs can also be computed as follows:
\begin{equation}
    k_{\alpha\beta\gamma}^{\mathrm{eq}} = \sum_i {
    f_i^{\mathrm{eq}} 
    (c_{ix} - u_x)^\alpha
    (c_{iy} - u_y)^\beta
    (c_{iz} - u_z)^\gamma}.
    \label{eq:definition_equilibrium_central_moments}
\end{equation}
Here, we investigate the equilibrium CMs for several equilibrium distribution functions.
Scripts with symbolic computations used to specifically compute the equilibrium CMs \texttt{get\_equilibria.jl} can be found in the Supplementary Material.

\subsection{Standard CG equilibria \texorpdfstring{with $g_i^{\mathrm{eq},2}$}{up to second order} \label{sec:standard_equilibria_up_to_2nd}}

We begin with the most popular equilibrium distribution function in the CG model, which is expressed in Eq.~(\ref{eq:standard_CG_2nd}).
By setting $N = 2$ and $\Phi_i = 0$ in Eq.~(\ref{eq:equilibria_in_general_form}), Eq.~(\ref{eq:standard_CG_2nd}) can be reformulated as:
\begin{equation}
    f_i^{\mathrm{eq}} = g_i^{\mathrm{eq},2}
        + (p - \rho c_s^2)E_i.
    \label{eq:CG_equilibria_with_standard_2nd_with_Ei} 
\end{equation}
Substituting Eq.~(\ref{eq:CG_equilibria_with_standard_2nd_with_Ei}) into Eq. ~(\ref{eq:definition_equilibrium_central_moments}), we compute the corresponding equilibrium CMs in each order as follows:

\noindent Zeroth order: 
\begin{equation}
    k_{000}^{\mathrm{eq}} =  \rho,
    \label{eq:equilibrium_central_moments_standard2nd_0th}
\end{equation}
First order:
\begin{equation}
  k_{100}^{\mathrm{eq}} = k_{010}^{\mathrm{eq}} = k_{001}^{\mathrm{eq}} = 0  , 
\end{equation}
Second order:
\begin{equation}
  \begin{split}
  k_{110}^{\mathrm{eq}} = k_{011}^{\mathrm{eq}} = k_{101}^{\mathrm{eq}} = 0  ,  \\
  k_{200}^{\mathrm{eq}} = k_{020}^{\mathrm{eq}} = k_{002}^{\mathrm{eq}} = p  ,  \\
  \end{split}
  \label{eq:equilibrium_central_moments_standard2nd_2nd}
\end{equation}
Third order:
\begin{equation}
  \begin{split}
  % 3rd order
  k_{120}^{\mathrm{eq}} = & \underline{-(p - \rho c_s^2)u_x}  \underline{\underline{{- \rho u_x u_y ^2}}}, \\
  k_{102}^{\mathrm{eq}} = & \underline{-(p - \rho c_s^2)u_x}  \underline{\underline{{- \rho u_x u_z ^2}}},  \\
  k_{012}^{\mathrm{eq}} = & \underline{-(p - \rho c_s^2)u_y}  \underline{\underline{{- \rho u_y u_z ^2}}}, \\
  k_{210}^{\mathrm{eq}} = & \underline{-(p - \rho c_s^2)u_y}  \underline{\underline{{- \rho u_x ^2u_y}}} ,  \\
  k_{201}^{\mathrm{eq}} = & \underline{-(p - \rho c_s^2)u_z}  \underline{\underline{{- \rho u_x ^2u_z}}} , \\
  k_{021}^{\mathrm{eq}} = & \underline{-(p - \rho c_s^2)u_z}  \underline{\underline{{- \rho u_y ^2u_z}}} ,  \\
  k_{111}^{\mathrm{eq}} = & \underline{\underline{{-\rho u_x  u_y  u_z}}} ,  \\
  \label{eq:equilibrium_central_moments_standard2nd_3rd}
  \end{split}
\end{equation}
Fourth order:
\begin{equation}
  \begin{split}
  % 4th order
  k_{220}^{\mathrm{eq}} = & ~ p c_s^2 \underline{+(p - \rho c_s^2)(u_x ^2 + u_y ^2)} \underline{\underline{{+ 3 \rho u_x ^2 u_y ^2}}} ,  \\
  k_{202}^{\mathrm{eq}} = & ~ p c_s^2 \underline{+(p - \rho c_s^2)(u_x ^2 + u_z ^2)} \underline{\underline{{+ 3 \rho u_x ^2 u_z ^2}}} ,  \\
  k_{022}^{\mathrm{eq}} = & ~ p c_s^2 \underline{+(p - \rho c_s^2)(u_y ^2 + u_z ^2)} \underline{\underline{{+ 3 \rho u_y ^2 u_z ^2}}} ,  \\
  k_{211}^{\mathrm{eq}} = & ~ \underline{(p - \rho c_s^2) u_y  u_z}  \underline{\underline{{+ 3 \rho u_x ^2 u_y  u_z}}} ,  \\
  k_{121}^{\mathrm{eq}} = & ~ \underline{(p - \rho c_s^2) u_x  u_z}  \underline{\underline{{+ 3 \rho u_x  u_y ^2 u_z}}} ,  \\
  k_{112}^{\mathrm{eq}} = & ~ \underline{(p - \rho c_s^2) u_x  u_y}  \underline{\underline{{+ 3 \rho u_x  u_y  u_z ^2}}},  \\
  \end{split}
\end{equation}
Fifth order:
\begin{equation}
  \begin{split}
  % Fifth order:
  k_{122}^{\mathrm{eq}} = & \underline{- (p - \rho c_s^2)u_x ( u_y ^2 + u_z ^2 + c_s^2)} \\
  & \underline{\underline{{- \rho u_x \qty[6  u_y ^2 u_z ^2 +  c_s^2 (u_y ^2 + u_z ^2)]}}} ,  \\
  k_{212}^{\mathrm{eq}} = & \underline{- (p - \rho c_s^2)u_y ( u_x ^2 + u_z ^2 + c_s^2)} \\
  & \underline{\underline{{- \rho u_y \qty[6 u_x ^2  u_z ^2 + c_s^2 (u_x ^2 + u_z ^2)]}}} ,  \\
  k_{221}^{\mathrm{eq}} = & \underline{- (p - \rho c_s^2)u_z ( u_x ^2 + u_y ^2 + c_s^2)} \\
  & \underline{\underline{{-  \rho u_z \qty[6 u_x ^2 u_y ^2  + c_s^2 (u_x ^2 + u_y ^2)]}}} ,  \\
  \end{split}
\end{equation}
Sixth order:
\begin{equation}
  \begin{split}
  % 6th order
  k_{222}^{\mathrm{eq}} = &  ~ p c_s^4 + \underline{( p - \rho c_s^2 )[u_x ^2 u_y ^2 + u_x ^2 u_z ^2 + u_y ^2 u_z ^2}   \\
    & \underline{+ c_s^2(u_x ^2 + u_y ^2 + u_z ^2)]} \\
    & \underline{\underline{{+ \rho \qty[10u_x ^2 u_y ^2 u_z ^2 + 3c_s^2(u_x ^2 u_y ^2 + u_x ^2 u_z ^2 + u_y ^2 u_z ^2)]}}}.
    \label{eq:equilibrium_central_moments_standard2nd_6th}
  \end{split}
\end{equation}
Based on Eqs.~(\ref{eq:equilibrium_central_moments_standard2nd_0th})--(\ref{eq:equilibrium_central_moments_standard2nd_6th}), several properties of the equilibrium CMs in Eq.~(\ref{eq:CG_equilibria_with_standard_2nd_with_Ei}) can be identified.
\begin{itemize}
    \item The underlined terms: products of $(p-\rho c_s^2)$ and $u_\alpha$.
    \item The double underlined terms: products of $\rho$ and $u_\alpha$.
    \item The remaining terms: independent of $u_\alpha$.
\end{itemize}

%%%%%%%%%%%%%%%%%% MOMENTS %%%%%%%%%%%%%%%%%%%%%%%
% Taking the third-order moments of the equilibrium distribution function given by Eq.~(\ref{eq:CG_equilibria_with_standard_2nd_with_Ei}), we can obtain
% \begin{align}
%     \sum_i f_i^{\mathrm{eq}} c_{i\alpha} c_{i\beta} &= \rho u_\alpha u_\beta + p \delta_{\alpha\beta}, \label{eq:second_order_moment_standard_up_to_2nd}  \\
%     \sum_i f_i^{\mathrm{eq}} c_{i\alpha} c_{i\beta} c_{i\gamma} &= \rho c_s^2 (u_\alpha \delta_{\beta\gamma} + u_\beta \delta_{\alpha\gamma} + u_\gamma \delta_{\alpha\beta}), \label{eq:third_order_moment_standard_up_to_2nd}
% \end{align}
%%%%%%%%%%%%%%%%%% MOMENTS %%%%%%%%%%%%%%%%%%%%%%%

\subsection{Standard CG equilibria \texorpdfstring{with $g_i^{\mathrm{eq},6}$}{up to sixth order}
\label{sec:standard_equilibria_up_to_6th}}

In single-phase flow models, upon applying the correct set of Hermite polynomials to the discrete equilibrium, the velocity dependence in the derived equilibrium CMs vanishes, leading to Galilean invariance~\citep{De_Rosis2019-bq}.
To demonstrate the effects of similar procedures on the CG model, 
we set $N=6$ and $\Phi_i = 0$ in Eq.~(\ref{eq:equilibria_in_general_form}):
\begin{equation}
      f_i^{\mathrm{eq}} = g_i^{\mathrm{eq},6}
    + (p - \rho c_s^2)E_i.
    \label{eq:CG_equilibria_with_standard_6th} 
\end{equation}
This procedure corresponds to the replacement of $g_i^{\mathrm{eq},2}$ in Eq.~(\ref{eq:CG_equilibria_with_standard_2nd_with_Ei}) using $g_i^{\mathrm{eq},6}$ [Eq.~(\ref{eq:equilibria_for_standard_LBM_6th})].
The equilibrium distribution function adopted in Ref.~\citep{De_Rosis2019-bi} is similar to Eq.~(\ref{eq:CG_equilibria_with_standard_6th}).
However, the previous study uses a different form of correction operator, unlike the method used in Refs.~\citep{Leclaire2013-nx,Saito2018-ub} [see Eq.~(D5) in Ref.~\citep{De_Rosis2019-bi} for details].
Substituting Eq.~(\ref{eq:CG_equilibria_with_standard_6th}) into Eq.~(\ref{eq:definition_equilibrium_central_moments}), the corresponding equilibrium CMs are computed as:

\noindent Zeroth order: 
\begin{equation}
    k_{000}^{\mathrm{eq}} =  \rho,
\end{equation}
First order:
\begin{equation}
  k_{100}^{\mathrm{eq}} = k_{010}^{\mathrm{eq}} = k_{001}^{\mathrm{eq}} = 0  , 
\end{equation}
Second order:
\begin{equation}
  \begin{split}
  k_{110}^{\mathrm{eq}} = k_{011}^{\mathrm{eq}} = k_{101}^{\mathrm{eq}} = 0  ,  \\
  k_{200}^{\mathrm{eq}} = k_{020}^{\mathrm{eq}} = k_{002}^{\mathrm{eq}} = p  ,  \\
  \end{split}
\end{equation}
Third order:
\begin{equation}
  \begin{split}
  % 3rd order
  &k_{120}^{\mathrm{eq}} = k_{102}^{\mathrm{eq}} = \underline{-(p - \rho c_s^2)u_x} , \\
  &k_{012}^{\mathrm{eq}} = k_{210}^{\mathrm{eq}} = \underline{-(p - \rho c_s^2)u_y} , \\
  &k_{201}^{\mathrm{eq}} = k_{021}^{\mathrm{eq}} = \underline{-(p - \rho c_s^2)u_z} , \\
  &k_{111}^{\mathrm{eq}} = 0, \\
  \label{eq:equilibrium_central_moments_standard6th_3rd}
  \end{split}
\end{equation}
Fourth order:
\begin{equation}
  \begin{split}
  % 4th order
  k_{220}^{\mathrm{eq}} = & ~ p c_s^2 \underline{+ (p - \rho c_s^2)(u_x ^2 + u_y ^2)},  \\
  k_{202}^{\mathrm{eq}} = & ~ p c_s^2 \underline{+ (p - \rho c_s^2)(u_x ^2 + u_z ^2)},  \\
  k_{022}^{\mathrm{eq}} = & ~ p c_s^2 \underline{+ (p - \rho c_s^2)(u_y ^2 + u_z ^2)},  \\
  k_{211}^{\mathrm{eq}} = & ~ \underline{(p - \rho c_s^2) u_y  u_z} , \\
  k_{121}^{\mathrm{eq}} = & ~ \underline{(p - \rho c_s^2) u_x  u_z} ,  \\
  k_{112}^{\mathrm{eq}} = & ~ \underline{(p - \rho c_s^2) u_x  u_y} ,  \\
  \label{eq:equilibrium_central_moments_standard6th_4th}
  \end{split}
\end{equation}
Fifth order: 
\begin{equation}
  \begin{split}
  % Fifth order:
  k_{122}^{\mathrm{eq}} = & \underline{- (p - \rho c_s^2)u_x ( u_y ^2 + u_z ^2 + c_s^2)},  \\
  k_{212}^{\mathrm{eq}} = & \underline{- (p - \rho c_s^2)u_y ( u_x ^2 + u_z ^2 + c_s^2)},  \\
  k_{221}^{\mathrm{eq}} = & \underline{- (p - \rho c_s^2)u_z ( u_x ^2 + u_y ^2 + c_s^2)},  \\
  \end{split}
\end{equation}
Sixth order:
\begin{equation}
  \begin{split}
  % 6th order
  k_{222}^{\mathrm{eq}} = &  ~ p c_s^4 \underline{+ ( p - \rho c_s^2 )
  [u_x ^2 u_y ^2 + u_x ^2 u_z ^2 + u_y ^2 u_z ^2}   \\
    & \underline{+ c_s^2(u_x ^2 + u_y ^2 + u_z ^2)]}.
    \label{eq:equilibrium_central_moments_standard6th_6th}
  \end{split}
\end{equation}
No change is observed in the equilibrium from the zeroth to second order; they are identical to those in Eqs.~(\ref{eq:equilibrium_central_moments_standard2nd_0th})--(\ref{eq:equilibrium_central_moments_standard2nd_2nd}) in Sec.~\ref{sec:standard_equilibria_up_to_2nd}.
This is because the equilibrium distribution functions in Eqs.~(\ref{eq:CG_equilibria_with_standard_2nd_with_Ei}) and (\ref{eq:CG_equilibria_with_standard_6th}) are identical from the zeroth to second order.
Upon comparing with Eqs.~(\ref{eq:equilibrium_central_moments_standard2nd_3rd})--(\ref{eq:equilibrium_central_moments_standard2nd_6th}), it can be seen that differences appear at the equilibrium CMs for the third and higher orders.
That is, the double-underlined terms in the form of the product of $\rho$ and the velocity components, as seen in Eqs.~(\ref{eq:equilibrium_central_moments_standard2nd_3rd})--(\ref{eq:equilibrium_central_moments_standard2nd_6th}), are now completely removed.
Therefore, one of the third-order moments $k_{111}^{\mathrm{eq}}$ becomes zero,
However, there are still underlined velocity-dependent terms proportional to $(p - \rho c_s^2)$, as shown in Eqs.~(\ref{eq:equilibrium_central_moments_standard6th_3rd})--(\ref{eq:equilibrium_central_moments_standard6th_6th}).

%%%%%%%%%%%%%%%%%% MOMENTS %%%%%%%%%%%%%%%%%%%%%%%
% \color{red}
% The second-order moment of Eq.~(\ref{eq:CG_equilibria_with_standard_6th}) is identical to Eq.~(\ref{eq:second_order_moment_standard_up_to_2nd}).
% The third-order moment of Eq.~(\ref{eq:CG_equilibria_with_standard_6th}), due to the contribution of the higher-order terms in $g_i^\mathrm{eq,6}$, reads
% % \begin{equation}
% % \begin{split}
% %         & \sum_i f_i^{\mathrm{eq}} c_{i\alpha} c_{i\beta} c_{i\gamma} \\
% %         & = 
% %         \begin{cases}
% %             \rho c_s^2 (u_\alpha \delta_{\beta\gamma} + u_\beta \delta_{\alpha\gamma} + u_\gamma \delta_{\alpha\beta}), & \mathrm{if}~\alpha = \beta = \gamma \\
% %             \rho c_s^2 (u_\alpha \delta_{\beta\gamma} + u_\beta \delta_{\alpha\gamma} + u_\gamma \delta_{\alpha\beta}) + \rho u_\alpha u_\beta u_\gamma, & \mathrm{others}
% %         \end{cases}
% % \end{split}
% % \end{equation}
% \begin{equation}
% \begin{split}
%         \sum_i f_i^{\mathrm{eq}} c_{i\alpha} c_{i\beta} c_{i\gamma}
%         &= \rho c_s^2 (u_\alpha \delta_{\beta\gamma} + u_\beta \delta_{\alpha\gamma} + u_\gamma \delta_{\alpha\beta}) \\
%         &\phantom{=} + \delta_{\alpha\beta} \delta_{\alpha\gamma} \rho u_\alpha u_\beta u_\gamma
% \end{split}
% \end{equation}
% \color{black}
%%%%%%%%%%%%%%%%%% MOMENTS %%%%%%%%%%%%%%%%%%%%%%%

\subsection{Improved equilibria by Li \textit{et al.}}

With the standard equilibrium distribution function in Eq.~(\ref{eq:standard_CG_2nd}) [or Eq.~(\ref{eq:CG_equilibria_with_standard_2nd_with_Ei})], the recovered macroscopic equation includes an unwanted error term~\citep{Liu2012-fa}, as argued by \citet{Huang2013-pd}.
To reduce the effect of this error term, Li \textit{et al.}~\citep{Ba2016-ve,Wen2019-jc} proposed an improved equilibrium distribution function based on their previous research~\citep{Li2012-vr}.
The original form of this improved equilibrium distribution function can be formulated as:
\begin{widetext}
% \begin{equation}
% \begin{split}
%   f_i^{\mathrm{eq}} &= \rho \biggl( \varphi_i + w_i \biggl[ \frac{\vb{c}_i \cdot \vb{u}}{c_s^2} + \frac{(\vb{c}_i \cdot \vb{u})^2}{2c_s^4} - \frac{|\vb{u}|^2}{2c_s^2} \\
%     & \qquad + \frac{\vb{c}_i \cdot \vb{u}}{2c_s^2} \qty(\frac{p}{\rho c_s^2} - 1)\qty(\frac{|\vb{c}_i|^2}{c_s^2} - D-2) \biggr] \biggr),
%     \label{eq:Li_and_coworkers_Equilibriua_original}
% \end{split}
% \end{equation}
\begin{equation}
  f_i^{\mathrm{eq}} = \rho \biggl( \varphi_i + w_i \biggl[ \frac{\vb{c}_i \cdot \vb{u}}{c_s^2} + \frac{(\vb{c}_i \cdot \vb{u})^2}{2c_s^4} - \frac{|\vb{u}|^2}{2c_s^2} 
     + \frac{\vb{c}_i \cdot \vb{u}}{2c_s^2} \qty(\frac{p}{\rho c_s^2} - 1)\qty(\frac{|\vb{c}_i|^2}{c_s^2} - D-2) \biggr] \biggr),
    \label{eq:Li_and_coworkers_Equilibriua_original}
\end{equation}
where $D$ denotes the spatial dimension.
The last term of Eq.~(\ref{eq:Li_and_coworkers_Equilibriua_original}) is derived on the basis of a third-order Hermite expansion of the Maxwell--Boltzmann distribution~\citep{Shan2006-tf,Li2012-vr}.
Upon rewriting Eq.~(\ref{eq:Li_and_coworkers_Equilibriua_original}) in a form equivalent to Eq.~(\ref{eq:equilibria_in_general_form}), the nonzero correction operator $\Phi_i$ is obtained as:
\begin{equation}
\begin{split}
  f_i^{\mathrm{eq}} &= g_i^{\mathrm{eq},2} 
  + (p-\rho c_s^2) \qty(E_i + \Phi_i) \\
   &= g_i^{\mathrm{eq},2} 
  + (p-\rho c_s^2) \qty(E_i + w_i
  \qty[\frac{u_x \qty(H_{i120} + H_{i102}) + u_y (H_{i210} + H_{i012}) + u_z (H_{i201} + H_{i021})}{2c_s^6} 
   ]).
   \label{eq:Li_and_coworkers_Equilibriua_in_general_form}
\end{split}
\end{equation}
\end{widetext}
% Taking the third-order moments of the equilibrium distribution function given by Eq.~(\ref{eq:Li_and_coworkers_Equilibriua_in_general_form}), we can obtain~\citep{Wen2019-jc}
% \begin{equation}
% \begin{split}
%         &\sum_i f_i^{\mathrm{eq}} c_{i\alpha} c_{i\beta} c_{i\gamma} \\
%         &= 
%         \begin{cases}
%             \rho c_s^2 (u_\alpha \delta_{\beta\gamma} + u_\beta \delta_{\alpha\gamma} + u_\gamma \delta_{\alpha\beta}), & \mathrm{if}~\alpha = \beta = \gamma \\
%             % p (u_\alpha \delta_{\beta\gamma} + u_\beta \delta_{\alpha\gamma} + u_\gamma \delta_{\alpha\beta}) + \rho u_\alpha u_\beta u_\gamma, & \mathrm{others}
%             p (u_\alpha \delta_{\beta\gamma} + u_\beta \delta_{\alpha\gamma} + u_\gamma \delta_{\alpha\beta}), & \mathrm{others}
%         \end{cases}
%         \label{eq:third_order_moment_of_Li_and_coworkers_equilibria}
% \end{split}
% \end{equation}
% Compared with Eq.~(\ref{eq:third_order_moment_standard_up_to_2nd}), $\rho c_s^2$  (the pressure defined in single-phase flow) in the off-diagonal elements of the third-order moment has been replaced by $p$ [pressure defined in the present multiphase flow model Eq.~(\ref{eq:pressure})];
% however, the diagonal elements still have $\rho c_s^2$ due to the low symmetry of standard lattices~\citep{Wen2019-jc}.
% We shall present procedures to correct this inconsistency later in~\ref{sec:Correction_for_the_diagonal_elements}.

Here, we replace $g_i^{\mathrm{eq},2}$ in Eq.~(\ref{eq:Li_and_coworkers_Equilibriua_in_general_form}) using $g_i^{\mathrm{eq},6}$, as described in Sec.~\ref{sec:standard_equilibria_up_to_6th} and then substituting it into Eq.~(\ref{eq:definition_equilibrium_central_moments}) to obtain the equilibrium CMs, which yields the following results:

\noindent Zeroth order: 
\begin{equation}
    k_{000}^{\mathrm{eq}} =  \rho,
    \label{eq:equilibrium_central_moments_improved3rd_0th}
\end{equation}
First order:
\begin{equation}
  k_{100}^{\mathrm{eq}} = k_{010}^{\mathrm{eq}} = k_{001}^{\mathrm{eq}} = 0  , 
\end{equation}
Second order:
\begin{equation}
  \begin{split}
  k_{110}^{\mathrm{eq}} = k_{011}^{\mathrm{eq}} = k_{101}^{\mathrm{eq}} = 0  ,  \\
  k_{200}^{\mathrm{eq}} = k_{020}^{\mathrm{eq}} = k_{002}^{\mathrm{eq}} = p  ,  \\
  \end{split}
  \label{eq:equilibrium_central_moments_improved3rd_2nd}
\end{equation}
Third order:
\begin{equation}
  \begin{split}
  % 3rd order
  k_{120}^{\mathrm{eq}} = k_{102}^{\mathrm{eq}} = k_{012}^{\mathrm{eq}} = k_{210}^{\mathrm{eq}} = k_{201}^{\mathrm{eq}} = k_{021}^{\mathrm{eq}} = k_{111}^{\mathrm{eq}} = 0 \label{eq:equilibrium_central_moments_improved3rd_3rd}
  \end{split}
\end{equation}
Fourth order:
\begin{equation}
  \begin{split}
  % 4th order
  k_{220}^{\mathrm{eq}} = & ~ p c_s^2 \underline{- (p - \rho c_s^2)(u_x ^2 + u_y ^2)},  \\
  k_{202}^{\mathrm{eq}} = & ~ p c_s^2 \underline{- (p - \rho c_s^2)(u_x ^2 + u_z ^2)},  \\
  k_{022}^{\mathrm{eq}} = & ~ p c_s^2 \underline{- (p - \rho c_s^2)(u_y ^2 + u_z ^2)},  \\
  k_{211}^{\mathrm{eq}} = & ~ \underline{-(p - \rho c_s^2) u_y  u_z} , \\
  k_{121}^{\mathrm{eq}} = & ~ \underline{-(p - \rho c_s^2) u_x  u_z} ,  \\
  k_{112}^{\mathrm{eq}} = & ~ \underline{-(p - \rho c_s^2) u_x  u_y} ,  \\
  \label{eq:equilibrium_central_moments_improved3rd_4th}
  \end{split}
\end{equation}
Fifth order:
\begin{equation}
  \begin{split}
  % Fifth order:
  k_{122}^{\mathrm{eq}} = & \underline{(p - \rho c_s^2)u_x ( 2u_y ^2 + 2u_z ^2 + c_s^2)},  \\
  k_{212}^{\mathrm{eq}} = & \underline{(p - \rho c_s^2)u_y ( 2u_x ^2 + 2u_z ^2 + c_s^2)},  \\
  k_{221}^{\mathrm{eq}} = & \underline{(p - \rho c_s^2)u_z ( 2u_x ^2 + 2u_y ^2 + c_s^2)},  \\
  \end{split}
\end{equation}
Sixth order:
\begin{equation}
  \begin{split}
  % 6th order
  k_{222}^{\mathrm{eq}} = &  ~ p c_s^4 \underline{- 3( p - \rho c_s^2 )( u_x ^2 u_y ^2 + u_x ^2 u_z ^2 + u_y ^2 u_z ^2}   \\
    & \underline{+ c_s^2(u_x ^2 + u_y ^2 + u_z ^2))},
    \label{eq:equilibrium_central_moments_improved3rd_6th}
  \end{split}
\end{equation}
The equilibrium CMs from the zeroth to second order remain unchanged; the higher-order terms above the fourth order change slightly compared to Eqs.~(\ref{eq:equilibrium_central_moments_standard6th_4th})--(\ref{eq:equilibrium_central_moments_standard6th_6th}). However, there still remain underlined velocity-dependent terms proportional to $(p-\rho c_s^2)$.
Notably, all of the third-order equilibrium moments in Eq.~(\ref{eq:equilibrium_central_moments_standard6th_3rd}) become zero.
By using the modified equilibrium distribution function in Eq.~(\ref{eq:Li_and_coworkers_Equilibriua_in_general_form}), based on the \textit{third-order} Hermite expansion, the velocity-dependent terms proportional to $(p - \rho c_s^2)$ in the \textit{third-order} equilibrium CMs can be eliminated; i.e., it has been implied that the $N$th-order velocity-dependent terms proportional to $(p-\rho c_s^2)$ in the equilibrium CMs can be eliminated by considering the appropriate form of $\Phi_i$ in Eq.~(\ref{eq:equilibria_in_general_form}) with the $N$th-order Hermite polynomials.

\subsection{Generalized equilibria}
Based on the above results and considerations, we propose a new generalized equilibrium distribution function for the CG model.
Similar to $g_i^{\mathrm{eq},N}$ in Eq.~(\ref{eq:equilibria_in_general_form}), we adopt an equilibrium distribution function based on Hermite expansion up to the sixth order $g_i^\mathrm{eq,6}$~\citep{De_Rosis2019-bq,De_Rosis2019-bi}.
In addition, considering the fourth-, fifth-, and sixth-order effects in $\Phi_i$ in Eq.~(\ref{eq:equilibria_in_general_form}), we define the following equilibrium distribution function:
\begin{widetext}
\begin{equation}
  \begin{split}
    f_i^{\mathrm{eq}} &= g_i^{\mathrm{eq},6} + 
      ( p - \rho c_s^2 ) 
      ( E_i + \Phi_i) \\
    &= g_i^{\mathrm{eq},6} + 
      ( p - \rho c_s^2 ) 
      \left( E_i + w_i \left[ 
      \frac{u_x (H_{i120} + H_{i102}) + u_y (H_{i210} + H_{i012}) + u_z (H_{i201} + H_{i021})}{2c_s^6} \right.\right. \\
       & \left.\left. + \frac{(u_x^2 + u_y^2)  H_{i220} + (u_y^2 + u_z^2)  H_{i022} + (u_x^2 + u_z^2)  H_{i202} + 2(u_y  u_z H_{i211} + u_x u_z H_{i121} + u_x u_y H_{i112})}{4c_s^8}   \right.\right.\\
       & \left.\left.  + \frac{u_x (u_y^2 + u_z^2 -c_s^2) H_{i122} + u_y (u_x^2 + u_z^2 -c_s^2) H_{i212} + u_z (u_x^2 + u_y^2 -c_s^2) H_{i221}}{4c_s^{10}}  \right.\right.\\
       & \left.\left.  + \frac{( u_x^2 u_y^2 + u_y^2 u_z^2 + u_x^2 u_z^2 -c_s^2(u_x^2 + u_y^2 + u_z^2) )H_{i222}}{8c_s^{12}}  \right] \right), 
       \label{eq:New_Equilibriua_in_general_form}
  \end{split}
  \end{equation}
\end{widetext}
Naturally, if we neglect the terms involving the Hermite polynomials of the fourth, fifth, and sixth orders and replace $g_i^\mathrm{eq,6}$ with $g_i^\mathrm{eq,2}$, Eq.~(\ref{eq:New_Equilibriua_in_general_form}) is reduced to the original equilibria, as proposed by Li \textit{et al.}~\citep{Ba2016-ve,Wen2019-jc} and described by Eq.~(\ref{eq:Li_and_coworkers_Equilibriua_original}) or (\ref{eq:Li_and_coworkers_Equilibriua_in_general_form}).

In the phase space, the new equilibrium distribution function proposed in Eq.~(\ref{eq:New_Equilibriua_in_general_form}) contains terms up to $\mathrm{O}(u^6)$, increasing its complexity compared to the conventional form in Eq.~(\ref{eq:standard_CG_2nd}), and its implementation is cumbersome.
However, computing the equilibrium moments using Eq.~(\ref{eq:definition_equilibrium_central_moments}) yields the following results:

\noindent Zeroth order: 
\begin{equation}
    k_{000}^{\mathrm{eq}} =  \rho,
    \label{eq:equilibrium_central_moments_proposed_0th}
\end{equation}
First order:
\begin{equation}
  k_{100}^{\mathrm{eq}} = k_{010}^{\mathrm{eq}} = k_{001}^{\mathrm{eq}} = 0  , 
\end{equation}
Second order:
\begin{equation}
  \begin{split}
  k_{110}^{\mathrm{eq}} = k_{011}^{\mathrm{eq}} = k_{101}^{\mathrm{eq}} &= 0  ,  \\
  k_{200}^{\mathrm{eq}} = k_{020}^{\mathrm{eq}} = k_{002}^{\mathrm{eq}} &= p  ,  \\
  \end{split}
\end{equation}
Third order:
\begin{equation}
  \begin{split}
  % 3rd order
  k_{120}^{\mathrm{eq}} = k_{102}^{\mathrm{eq}} = k_{012}^{\mathrm{eq}} = k_{210}^{\mathrm{eq}} = k_{201}^{\mathrm{eq}} = k_{021}^{\mathrm{eq}} = k_{111}^{\mathrm{eq}} = 0, \label{eq:equilibrium_central_moments_proposed_3rd}
  \end{split}
\end{equation}
Fourth order:
\begin{equation}
  \begin{split}
  % 4th order
  k_{220}^{\mathrm{eq}} = k_{202}^{\mathrm{eq}} = k_{022}^{\mathrm{eq}} &=  p c_s^2,  \\
  k_{211}^{\mathrm{eq}} = k_{121}^{\mathrm{eq}} = k_{112}^{\mathrm{eq}} &= 0, \\
  \end{split}
\end{equation}
Fifth order:
\begin{equation}
  % Fifth order:
  k_{122}^{\mathrm{eq}} = k_{212}^{\mathrm{eq}} = k_{221}^{\mathrm{eq}} = 0, 
\end{equation}
Sixth order:
\begin{equation}
  % 6th order
  k_{222}^{\mathrm{eq}} =  p c_s^4.
  \label{eq:equilibrium_central_moments_proposed_6th}
\end{equation}
The equilibrium moments of the generalized equilibrium distribution function [Eq.~(\ref{eq:New_Equilibriua_in_general_form})] are no longer velocity-dependent in the CM space.
Nonzero equilibrium moments are limited to even orders, with only six such moments; their form is simple.
All the odd moments are zero.
For $p = \rho c_s^2$ as in the LB method for single-phase flows, this is attributed to the equilibrium CMs, as described in Refs.~\citep{Geier2006-jt,Premnath2011-ek,Geier2015-pv,Fei2018-ak,De_Rosis2019-bq}.
% \textcolor{red}{The equilibrium in the CM space is extremely simple; thus, its execution has tremendous advantages in terms of the computational cost and simplicity of coding.}
The equilibrium in the CM space is extremely simple; thus, its execution has also advantages in terms of coding simplicity.

\subsection{Correction term and implementation \label{sec:Correction_for_the_diagonal_elements}}

Considering the third-order moments of the equilibrium distribution function given by Eq.~(\ref{eq:New_Equilibriua_in_general_form}), we obtain:
\begin{equation}
\begin{split}
    &\sum_i f_i^{\mathrm{eq}} c_{i\alpha} c_{i\beta} c_{i\gamma} \\
    &= 
        \begin{cases}
            \rho c_s^2 (u_\alpha \delta_{\beta\gamma} + u_\beta \delta_{\alpha\gamma} + u_\gamma \delta_{\alpha\beta}), & \mathrm{if}~\alpha = \beta = \gamma \\
            p (u_\alpha \delta_{\beta\gamma} + u_\beta \delta_{\alpha\gamma} + u_\gamma \delta_{\alpha\beta}) + \rho u_\alpha u_\beta u_\gamma, & \mathrm{otherwise}
        \end{cases}
        \label{eq:third_order_moment_of_generalized_equilibria}
\end{split}
\end{equation}
The third-order moments of the proposed equilibrium distribution function are equivalent to those in Ref.~\citep{Wen2019-jc} except for the third-order term $\rho u_\alpha u_\beta u_\gamma$. 
As discussed in literature, even with corrections made to the equilibrium distribution function, the diagonal elements of the third-order moments contain the term $\rho c_s^2$ because of the low symmetry of the standard lattices.
To address this issue, a correction term $\vb{Q} = [Q_x,Q_y,Q_z]$ is computed and considered in the collision process, as described below~\citep{Ba2016-ve,Wen2019-jc}:
\begin{equation}
    \mathbf{Q} = -3 \nabla \cdot [(p - \rho c_s^2) \vb{u}].
    \label{eq:correction_term}
\end{equation}
The divergence operation is approximated using Eq.~(\ref{eq:second-order_isotropic_finite_difference}).
% \begin{equation}
% \begin{split}
%     Q_x = -3 \partial_x[ (p - \rho c_s^2 ) u_x ], \\
%     Q_y = -3 \partial_y[ (p - \rho c_s^2 ) u_y ], \\
%     Q_z = -3 \partial_z[ (p - \rho c_s^2 ) u_z ],
% \end{split}
% \end{equation}
The correction term computed using Eq.~(\ref{eq:correction_term}) is considered in the collision step as the external force on the relevant second-order diagonal moments ($k_{200}$, $k_{020}$, and $k_{002}$), as shown in Eq.~(\ref{eq:collision_step_in_general_form}).

In this study, all relaxation coefficients except for $\omega_1$ are set to unity as in the literature (e.g., Ref.~\citep{De_Rosis2019-bi}).
Regarding the bulk viscosity, this setting is considered to enhance the numerical stability of the present simulations by over-dissipating acoustic waves, as discussed in Ref.~\citep{Coreixas2019-ce}.
With such a choice of relaxation coefficients, the collision step in Eq.~(\ref{eq:collision_step_in_general_form}) can be reformulated as follows: 
\begin{equation}
  \begin{split}
  % 0th and 2nd order
  % k_{000}^{\mathrm{eq}} = &~ \rho, \\
  k_{100}^{*} &= F_x/2,  \\
  k_{010}^{*} &= F_y/2,  \\
  k_{001}^{*} &= F_z/2,  \\
  k_{110}^{*} &= (1 - \omega_1)k_{110},  \\
  k_{011}^{*} &= (1 - \omega_1)k_{011},  \\
  k_{101}^{*} &= (1 - \omega_1)k_{101}  ,  \\
  k_{200}^{*} - k_{020}^{*} &= (1 - \omega_1)(k_{200} - k_{020}) \\
  &+ (1 - \omega_1/2)(Q_x - Q_y),  \\
  k_{200}^{*} - k_{002}^{*} &= (1 - \omega_1)(k_{200} - k_{002}) \\
  &+ (1 - \omega_1/2)(Q_x - Q_z),  \\
  k_{200}^{*} + k_{020}^{*} + k_{002}^{*} &= 3p + (Q_x + Q_y + Q_z)/2,  \\
  % 3rd order
  k_{120}^{*} &= k_{120}^\mathrm{eq} + c_s^2 F_x/2, \\
  k_{102}^{*} &= k_{102}^\mathrm{eq} + c_s^2 F_x/2, \\
  k_{210}^{*} &= k_{210}^\mathrm{eq} + c_s^2 F_y/2, \\
  k_{012}^{*} &= k_{012}^\mathrm{eq} + c_s^2 F_y/2, \\
  k_{201}^{*} &= k_{201}^\mathrm{eq} + c_s^2 F_z/2, \\
  k_{021}^{*} &= k_{021}^\mathrm{eq} + c_s^2 F_z/2, \\
  k_{111}^{*} &= k_{111}^\mathrm{eq}, \\
  % 4th order
  k_{220}^{*} &= k_{220}^\mathrm{eq},  \\
  k_{202}^{*} &= k_{202}^\mathrm{eq},  \\
  k_{022}^{*} &= k_{022}^\mathrm{eq},  \\
  k_{211}^{*} &= k_{211}^\mathrm{eq}, \\
  k_{121}^{*} &= k_{121}^\mathrm{eq}, \\
  k_{112}^{*} &= k_{112}^\mathrm{eq}, \\
  % Fifth order:
  k_{122}^{*} &= k_{122}^\mathrm{eq} + c_s^4 F_x/2, \\
  k_{212}^{*} &= k_{212}^\mathrm{eq} + c_s^4 F_y/2, \\
  k_{221}^{*} &= k_{221}^\mathrm{eq} + c_s^4 F_z/2, \\
  % 6th order
  k_{222}^{*} &= k_{222}^\mathrm{eq}. \label{eq:collision_step_with_all_one}
  \end{split}
\end{equation}
Here, since our results so far have shown that $k_{100}^{\mathrm{eq}} = k_{010}^{\mathrm{eq}} = k_{001}^{\mathrm{eq}} = k_{110}^{\mathrm{eq}} = k_{011}^{\mathrm{eq}} = k_{101}^{\mathrm{eq}} = 0$ and $k_{200}^{\mathrm{eq}} = k_{020}^{\mathrm{eq}} = k_{002}^{\mathrm{eq}} = p$ for all types of equilibria, we have substituted these values.
When using the generalized equilibria proposed in this study [Eqs.~(\ref{eq:equilibrium_central_moments_proposed_3rd})--(\ref{eq:equilibrium_central_moments_proposed_6th})], the collision operations over the third order in Eq.~(\ref{eq:collision_step_with_all_one}) can be expressed more concisely:
\begin{equation}
    \begin{split}
  % % 0th and 2nd order
  % % k_{000}^{\mathrm{eq}} = &~ \rho, \\
  % k_{100}^{*} &= F_x/2,  \\
  % k_{010}^{*} &= F_y/2,  \\
  % k_{001}^{*} &= F_z/2,  \\
  % k_{110}^{*} &= (1 - \omega_1)k_{110},  \\
  % k_{011}^{*} &= (1 - \omega_1)k_{011},  \\
  % k_{101}^{*} &= (1 - \omega_1)k_{101}  ,  \\
  % k_{200}^{*} - k_{020}^{*} &= (1 - \omega_1)(k_{200} - k_{020}) \nonumber \\
  % & + (1 - \omega_1/2)(Q_x - Q_y),  \\
  % k_{200}^{*} - k_{002}^{*} &= (1 - \omega_1)(k_{200} - k_{002}) \nonumber \\
  % & + (1 - \omega_1/2)(Q_x - Q_z),  \\
  % k_{200}^{*} + k_{020}^{*} + k_{002}^{*} &= 3p + (Q_x + Q_y + Q_z)/2,  \\
  % 3rd order
  k_{120}^{*} = k_{102}^{*} &= c_s^2 F_x/2, \\
  k_{210}^{*} = k_{012}^{*} &= c_s^2 F_y/2, \\
  k_{201}^{*} = k_{021}^{*} &= c_s^2 F_z/2, \\
  k_{111}^{*} &= 0, \\
  % 4th order
  k_{220}^{*} = k_{202}^{*} = k_{022}^{*} &= p c_s^2,  \\
  k_{211}^{*} = k_{121}^{*} = k_{112}^{*} &= 0, \\
  % Fifth order:
  k_{122}^{*} &= c_s^4 F_x/2, \\
  k_{212}^{*} &= c_s^4 F_y/2, \\
  k_{221}^{*} &= c_s^4 F_z/2, \\
  % 6th order
  k_{222}^{*} &= p c_s^4, 
  \end{split}
\end{equation}
This implies that only six second-order CMs ($k_{110}$, $k_{011}$, $k_{101}$, $k_{200}$, $k_{020}$, and $k_{002}$) must be computed immediately before the collision operation, and the other CMs need not be computed in this case.
This not only simplifies the calculations but is also expected to reduce the computational cost.

% \subsection{Summary of equilibria}

The equilibria models considered in this study are summarized in Table~\ref{tab:summari_of_equilibria}.

\begin{table*}[tb]
\caption{
  \label{tab:summari_of_equilibria}
  Summary of equilibria models considered in this study.
  }
  \begin{ruledtabular}
    \begin{tabular}{lccc}
              &    Maximum order $N$ in $g_i^{\mathrm{eq},N}$  & Correction operator $\Phi_i$  & Consideration of $\vb{Q}$  \\
    \hline
      Model A  &  2 & 0       & No        \\
      Model B &  6     & 0        & No       \\
      Model C &  6      & Up to $O(u^3)$        & Yes   \\
      Model D (proposed in this paper) &  6  & Up to $O(u^6)$        & Yes   \\
    \end{tabular}
  \end{ruledtabular}
\end{table*}

\section{Numerical results and discussion\label{sec:numerical_experiments}}

To investigate the numerical properties of the generalized central equilibrium moments obtained in Sec.~\ref{sec:equilibria}, five numerical experiments are performed.
Comparisons with existing equilibrium distribution functions are also presented.
To ensure a fair comparison, all collision operations are performed in the CM space.
All relaxation coefficients are set to 1, except those related to the kinematic viscosity.
As wall boundaries in this paper, we implement the no-slip, slip, and moving wall boundary conditions using the fullway bounce-back scheme~\citep{Kruger2017-ux} for simplicity.
The computation code was written in the \texttt{Julia} language~\cite{Bezanson2017-jc}, version 1.6.7, and parallelized using the \texttt{MPI.jl} package ~\cite{Byrne2021-nq}.

\subsection{Stationary droplet}
The first case involves the simulation of a stationary droplet.
As a fundamental property of the numerical performance involving fluid--fluid interfaces, the validity of the surface tension obtained using the CG model described is checked, as described in Sec.~\ref{sec:color-gradient_model}.
By measuring the pressure difference between the inside and outside of the droplet at equilibrium, the surface tension is expressed, as shown in Eq.~(\ref{eq:CSF_force}).
It should be emphasized that, in principle, the CG model does not require prior computations to obtain the surface tension coefficient. 
However, this computation was performed for validation purposes.
Simulations were performed using Model D (Table~\ref{tab:summari_of_equilibria}).

Because the simulation was performed in two dimensions, according to Laplace's law, the pressures inside and outside the droplet were theoretically predicted as:
\begin{equation}
    \sigma = R \Delta p.
\end{equation}
The computational setup was adapted from Ref.~\citep{Ba2016-ve}.
A circular droplet of red fluid with a radius $R=25$ was placed in a $100 \times 100$ discretized space.
The surrounding area was filled with a blue fluid with $\rho_b^0=1$.
Periodic boundary conditions were imposed at all boundaries.
The kinematic viscosities of each phase were set to $\nu_r = \nu_b = 1/6$.
Other parameters were also set as in \citet{Ba2016-ve} and compared with the previous results.

Table~\ref{tab:Laplace_pressure} summarizes the simulation parameters and the error evaluations for certain density ratios and surface tension coefficients.
For stationary cases, the simulations are stable for density ratios up to $1000$.
The errors in the table are calculated as $E = |\sigma_\mathrm{th} - \sigma_\mathrm{cal}| / \sigma_\mathrm{th} \times 100\% $, where
the subscripts `th' and `cal' denote the theoretical and calculated surface tension coefficients, respectively.
By comparing the results of the existing RM-MRT model~\citep{Ba2016-ve} with those of the present CM-MRT model used in this study, we find that the evaluated errors are of the same order of magnitude.
Therefore, we can state that our generalized equilibria [Eqs.~(\ref{eq:equilibrium_central_moments_proposed_0th})--(\ref{eq:equilibrium_central_moments_proposed_6th})] can accurately predict the surface tension, which is pivotal for the simulation of two-phase flows.

\begin{table}[tb]
\caption{
  \label{tab:Laplace_pressure}
  Parameters used in the stationary droplet test and evaluated errors.
  The subscripts `th' and `cal' denote the theoretical and calculated surface tension coefficients, respectively.
  }
  \begin{ruledtabular}
    \begin{tabular}{ccccc}
    $\rho_r^0/\rho_b^0$ &  $\sigma_{\mathrm{th}}$  & $\sigma_{\mathrm{cal}}$  & $E$ \% & $E$ \%~\citep{Ba2016-ve} \\
    \hline
      1       & 0.012      & 0.01204       & 0.30  & 0.65     \\
      1       & 0.072     & 0.07223        & 0.31  & 0.84    \\
      1       & 0.24      & 0.2408        & 0.34   & 1.38  \\
      2       & 0.112      & 0.1122        & 0.21  & 0.95   \\
      10       & 0.132      & 0.1321        & 0.10  & 0.97  \\
      30       & 0.1116      & 0.11177        & 0.15 & 0.83  \\
      100       & 0.10      & 0.10032        & 0.32  & 0.69 \\
      1000       & 0.10      & 0.10089        & 0.89  & 0.74  \\
    \end{tabular}
  \end{ruledtabular}
\end{table}

\subsection{Droplet in a moving tube\label{sec:moving_droplet}}
Modifications to the standard equilibrium distribution function cause a lack of Galilean invariance in some free-energy~\citep{Inamuro2000-ol,Kalarakis2002-vm,Li2021-ar} and CG models~\citep{Wen2019-jc}.
As reported previously, if the Galilean invariance is lacking, the initially circular droplet in the moving tube is largely deformed.
In this section, we investigate the improvement in the equilibrium distribution function and the effect of the correction term described in Sec.~\ref{sec:Correction_for_the_diagonal_elements} for the recovery of the Galilean invariance.
We investigate the behavior of the four types of equilibria mentioned in Table~\ref{tab:summari_of_equilibria}.

We employ a computational setup similar to that in \citet{Wen2019-jc}.
The computational domain was discretized as $140\times140$ in a two-dimensional space.
Initially, a droplet with radius $R=30$ and density $\rho_r^0 = 3$ was placed at rest.
The droplet was surrounded by a stationary fluid with density $\rho_b^0=1$.
The top and bottom boundaries enforce a wall boundary condition moving at a constant velocity~\citep{Zou1997-pq,Zong2021-ne} with $U=0.02$, whereas periodic boundary conditions are set for the left and right boundaries.
The movements of the top and bottom walls drive the motion of the fluid inside, which in turn moves the droplets. 
The additional simulation parameters are as follows: $\beta=0.7$, $\mu_r = \mu_b = 0.075$, $\sigma = 4.4\times10^{-3}$.

\begin{figure}[tb]
    \centering
    \includegraphics[width=1\linewidth]{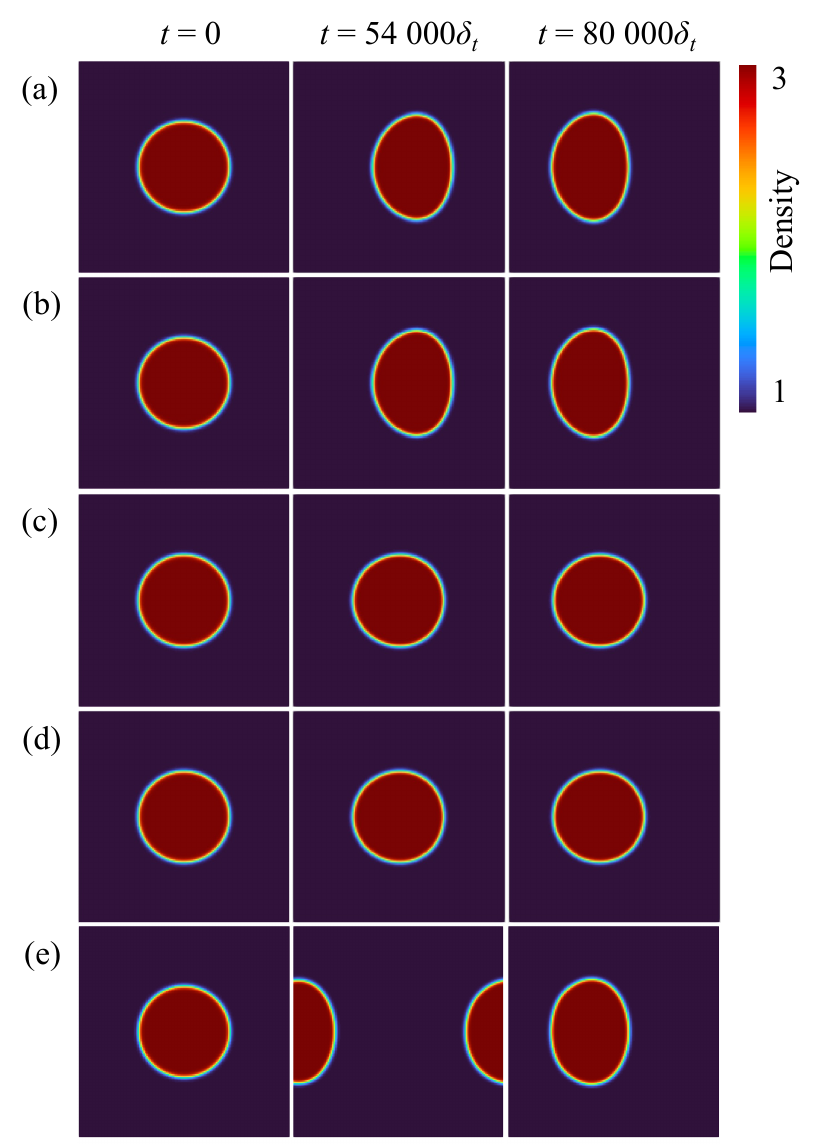}
    \caption{Density field of droplet in a moving tube: 
    (a) Model A, 
    (b) Model B, 
    (c) Model C, 
    (d) Model D. 
    Initially ($t=0$), a circular droplet is placed in a domain.
    Owing to the movement of the top and bottom walls, the droplet also begins to move.
    In Models A and B, the droplets transition to an elliptic shape over time owing to the lack of Galilean invariance.
    In Models C and D, the droplets remain nearly circular over time owing to improved Galilean invariance.
    To clarify whether $\Phi_i$ or $\mathbf{Q}$ contributes to the improvement in Galilean invariance, (e) an additional simulation based on Model D but without $\mathbf{Q}$, is also conducted. 
    In this case, the droplet still deforms over time.}
    \label{fig:moving_droplet_density_field}
\end{figure}

Images of the density distribution obtained using the conditions listed in Table~\ref{tab:summari_of_equilibria} are shown in Fig.~\ref{fig:moving_droplet_density_field}.
As shown in Figs.~\ref{fig:moving_droplet_density_field}(a) and (b), the droplets in both cases transition into elliptical shapes over time.
This is because of the lack of Galilean invariance in the equilibria employed in these models.
In the CG model, only considering $g_i^\mathrm{eq}$ up to the sixth order~\citep{De_Rosis2019-bq} does not inherently improve the Galilean invariance.
In contrast, Figs.~\ref{fig:moving_droplet_density_field}(c) and (d) show nearly equivalent and better results, respectively, than those in Figs.~\ref{fig:moving_droplet_density_field}(a) and (b). 
In other words, the droplets maintain their circular shape even after a long period of time.
Thus, we can see that Galilean invariance is restored in these models.
This is attributed to the contribution of the improvement in the third-order moments.

To clarify whether $\Phi_i$ or $\mathbf{Q}$ contributes to the improvement in Galilean invariance, an additional simulation based on Model D but without $\mathbf{Q}$ is also conducted [see Fig.~\ref{fig:moving_droplet_density_field}(e)].  
Although the shape and motion of the droplet are different from those in Figs.~\ref{fig:moving_droplet_density_field}(a) and (b), the droplet is deformed over time. 
Therefore, we conclude that \textit{both} the correction operator $\Phi_i$ of third order or higher and the correction term $\mathbf{Q}$ are necessary to improve the Galilean invariance of the CG model.

However, looking at again the results in Figs.~\ref{fig:moving_droplet_density_field}(c) and (d), one can see that the droplet is not perfectly circular, e.g., for $t = 54\,000$; therefore, it seems that the Galilean invariance is not \textit{perfectly} restored.
The gradient computation of the correction term $\mathbf{Q}$ [Eq.~(\ref{eq:correction_term})] by finite differences introduces numerical errors that distort the droplet.
A higher-order lattice (e.g., D3Q39 lattice) and the corresponding third-order equilibrium~\citep{Shan2006-tf} may be needed to solve it completely.
Interestingly,  no significant difference is observed between Figs.~\ref{fig:moving_droplet_density_field}(c) and (d).
This implies that at the Navier--Stokes level, the correction of the error term in the CG model is sufficient up to the third order in terms of accuracy.
This was to be expected because moments up to the fourth order are sufficient to recover even the compressible Navier--Stokes--Fourier equations~\citep{Shan2006-tf,Latt2020-cq}.

\begin{figure}[tb]
    \centering
    \includegraphics[width=1\linewidth]{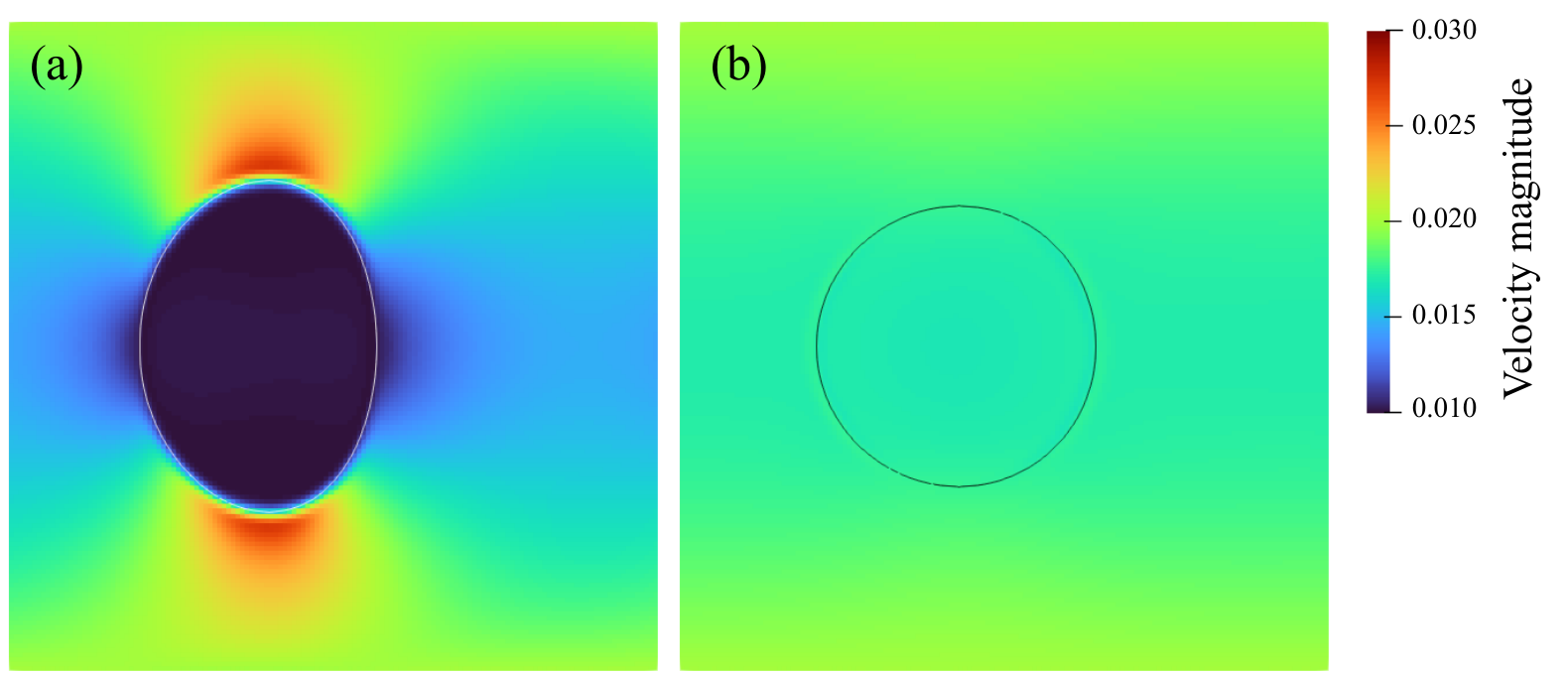}
    \caption{Comparison of the velocity field at $t=80\,000 \delta_t$: (a) Model A and (b) Model D. 
    The solid line in the figures represents the interface position.
    The correction of Galilean invariance significantly improves the discontinuity in the velocity distribution.}
    \label{fig:moving_droplet_velocity_field}
\end{figure}

To observe the differences in more detail, the velocity distribution characteristics in Models A and D are illustrated in Fig.~\ref{fig:moving_droplet_velocity_field}.
In Fig.~\ref{fig:moving_droplet_velocity_field}(a), an unphysical velocity distribution is observed around the deformed droplet.
A similar conclusion was drawn in a previous study~\citep{Ba2016-ve}.
However, as shown in Fig.~\ref{fig:moving_droplet_velocity_field}(b), no significant velocity discontinuity exists near the interface, and the velocity distribution is smooth.
This improvement in the unphysical flow is expected to result in a significant difference in the computational accuracy, especially for more complex computation targets.

\subsection{Layered two-phase flow\label{sec:layered-two-phase}}
To further investigate the effects of the four types of equilibria presented in Table~\ref{tab:summari_of_equilibria} on the velocity profiles of the two-phase flow, immiscible layered flows between two parallel plates~\citep{Huang2013-pd,Huang2015-fg} were simulated.
In the two-dimensional simulation, periodic boundary conditions were applied to the left and right boundaries, whereas no-slip boundaries were applied to the top and bottom boundaries.
A constant body force $\vb{F} = [F_x, 0]$ was applied to the entire domain as the driving force.
In this flow problem, the vertical velocity component $u_y$ is assumed to be zero throughout the domain.

Assuming a Poiseuille-type flow in the channel, the analytical solution for the velocity profile is given by~\citep{Huang2015-fg}:
\begin{equation} 
  u_x(y) =
      \begin{cases}
          A_1 y^2 + C_1,  & 0 \leq |y| \leq a, \\
          A_2 y^2 + B_2 y + C_2, & a \leq |y| \leq b, \\
      \end{cases}
      \label{eq:layered_two_phase_analytical_1}
\end{equation}
where the coefficients are defined as:
\begin{equation}
  \begin{split}
    A_1 = & -\frac{G}{2\mu_r}, ~A_2 = -\frac{G}{2\mu_b}, \\
    B_2 = & ~2(A_1 M - A_2)a, \\
    C_1 = & ~(A_2 - A_1) a^2 - B_2 (b-a) - A_2 b^2, \\
    C_2 = & -A_2 b^2 - B_2 b,
  \end{split}
  \label{eq:layered_two_phase_analytical_2}
\end{equation}
where $M = \mu_r/\mu_b$ denotes the dynamic shear-viscosity ratio.

In our simulations, the computational domain was discretized into a pseudo-one-dimensional setup with $N_x\times N_y=10\times 100$ lattices, where $a=25$ and $b=50$.
Three conditions equivalent to those in Ref.~\citep{Wen2019-jc} are considered, as summarized in Table~\ref{tab:layered_two_phase}, where the maximum density ratio reaches $1\,000$.
The surface tension coefficient and constant body force are set as $\sigma=0.002/4.5$ and $F_x=1.5 \times 10^{-8}$, respectively, as in \citet{Wen2019-jc}.
As described in Sec.~\ref{sec:moving_droplet}, the four equilibria models (Table~\ref{tab:summari_of_equilibria}) are examined.

\begin{table}[htbp]
\caption{
  \label{tab:layered_two_phase}
  Parameters used in the simulation of two-phase layered flow~\citep{Wen2019-jc}.
  }
  \begin{ruledtabular}
    \begin{tabular}{ccccc}
             &  $\rho_r^0$  & $\rho_b^0$  & $\alpha_b$ &  $M$ \\
    \hline
      Case A & 0.1      & 0.8       & 0.9    &   1/8 \\
      Case B & 0.8     & 0.1        & 0.2    &  8 \\
      Case C & 0.008      & 8       & 0.9992  & 1/40 \\
    \end{tabular}
  \end{ruledtabular}
\end{table}

\begin{figure*}
    \centering
    \includegraphics[width=1\linewidth]{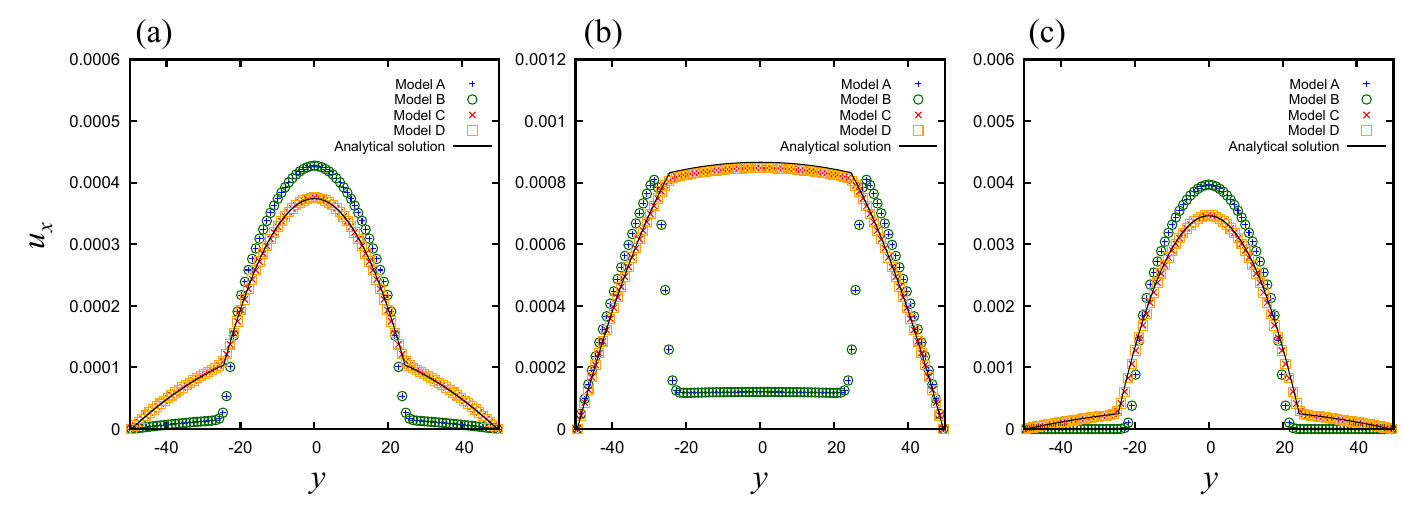}
    \caption{Velocity profiles of the simulated layered two-phase flow in a channel together with the analytical solution: (a) Case A, (b) Case B, and (c) Case C, as presented in Table~\ref{tab:layered_two_phase}.
    The four types of equilibria (Table~\ref{tab:summari_of_equilibria}) are examined.
    When a correction is applied (Models C and D), the velocity profiles obtained from the simulations agree well with the analytical solution.
    }
    \label{fig:layeredTwoPhase}
\end{figure*}

Figure~\ref{fig:layeredTwoPhase} shows the numerically obtained velocity profiles for each equilibrium presented in Table~\ref{tab:summari_of_equilibria} together with the analytical solution of Eq.~(\ref{eq:layered_two_phase_analytical_1}).
Similar to the simulation results described in Sec.~\ref{sec:moving_droplet}, Models A and B and Models C and D show nearly equivalent results.
Without correcting for the equilibria (Models A and B), the deviation between the numerical and analytical solutions increases, and a discontinuity near the interface can be observed in all three cases.
In contrast, when a correction term is applied (Models C and D), the velocity profiles obtained from the simulations agree well with the analytical solution.
Again, the accuracy at the Navier--Stokes level of Models C and D is comparable, as described in the previous section.

\subsection{Two-dimensional bubble rising}

To verify the accuracy of the \textit{dynamic} gas--liquid two-phase system, 
we performed a well-known single-bubble rising benchmark simulation, based on a previous study \citet{Hysing2009-cd}.
This is an unsteady problem involving parameters such as the density ratio, viscosity ratio, surface tension, interface curvature, and gravity; it is more complex than the simulations described above.
Because no exact solution to this problem can be obtained, a comparison with the available numerical solutions is made.
The dimensionless numbers describing the problem are the density ratio $\rho_r/\rho_b$, viscosity ratio $\mu_r/\mu_b$, Reynolds number $Re$, and Eotvos number $Eo$.
The Reynolds and Eotvos numbers are defined as follows:
\begin{align}
    Re = \frac{\rho_r U D}{\mu_r}, \\
    Eo = \frac{\rho_r U^2 D}{\sigma},
\end{align}
where $D$ denotes the bubble diameter (characteristic length scale), $U = (gD)^{1/2}$ is the characteristic velocity, and $\sigma$ is the surface tension coefficient.
The time scale is characterized by $D/U = (D/g)^{1/2}$. 
Here, we performed simulations for the two cases presented in Ref.~\citep{Hysing2009-cd}).
The cases and corresponding parameters in lattice units and dimensionless numbers are summarized in Table~\ref{tab:2dBubbleRisingParameters}.
For details on the computational setup and parameters, refer to Ref.~\citep{Hysing2009-cd}.

\begin{table*}[htbp]
\caption{
  \label{tab:2dBubbleRisingParameters}
  Physical parameters in lattice units and dimensionless numbers defining the test cases.
  }
  \begin{ruledtabular}
    \begin{tabular}{lcccccccccc}
     & $\rho_r^0$  & $\rho_b^0$  & $\mu_r$ & $\mu_b$ & $g$   & $\sigma$  & $Re$  & $Eo$  & $\rho_r^0/\rho_b^0$ & $\mu_r/\mu_b$ \\
    \hline
      Case 1 & 1.0 & 0.1   & $2.29\times 10^{-3}$  & $2.29\times 10^{-4}$ & $1.25\times 10^{-8}$  & $8.0\times 10^{-6}$  & 35    & 10    & 10              & 10   \\
      Case 2 & 1.0 & 0.001 & $4.57\times 10^{-3}$  & $4.57\times 10^{-5}$ & $6.25\times 10^{-9}$  & $1.28\times 10^{-6}$ & 35    & 125   & 1000            & 100 \\
    \end{tabular}
  \end{ruledtabular}
\end{table*}

The simulation was performed in a rectangular computational domain with width $W=1$ and height $H=2W$ in dimensionless units.
A bubble of diameter $D=0.5$ was placed at the center of the lower half of the domain in a domain filled with liquid.
This single bubble rose with gravity, with $\vb{F} = -\rho \vb{g}$ as the driving force.
No-slip boundary conditions were applied for the top and bottom edges, whereas slip boundary conditions were applied for the lateral sides.
Throughout this simulation, the relation $\alpha_b = 8/27$ is used to set the speed of sound in the gas phase to $c_s^b = 1/\sqrt{3}$, which is equivalent to that of the standard LB method using D3Q27~\citep{Saito2017-lg}.
In addition to the bubble shapes obtained from the simulation, according to ~\citep{Leclaire2016-oa}, the centroid position $y_c$ and bubble velocity $V_c$ are measured using the following definitions:
\begin{align}
    y_c = & \sum_{\vb{x}}{\rho_b(\vb{x}) y(\vb{x})} / \sum_{\vb{x}}{\rho_b(\vb{x})},\\
    V_c = & \sum_{\vb{x}}{\rho_b(\vb{x}) u_y(\vb{x})} / \sum_{\vb{x}}{\rho_b(\vb{x})} \label{eq:bubble_rise_velocity},
\end{align}
where the blue density $\rho_b(\vb{x})$ is used to track the bubble regions.
In recent years, this benchmark simulation has also been addressed by many researchers in the framework of the LB method to verify the accuracy of interface tracking~\citep{Sitompul2019-nr,Hajabdollahi2021-pv,Xu2021-hh,Baroudi2021-oi,Reis2022-lo}; it is noted that most of them are based on the phase-field LB model.
\citet{Leclaire2016-oa} was the first to address this benchmark problem in the CG model; however, in their simulation, only Case 1 was addressed due to the limitation of the density ratio.
Here, simulations are performed using only Model D, the generalized equilibria (Table~\ref{tab:summari_of_equilibria}).

\begin{figure*}
    \centering
    \includegraphics[width=1\linewidth]{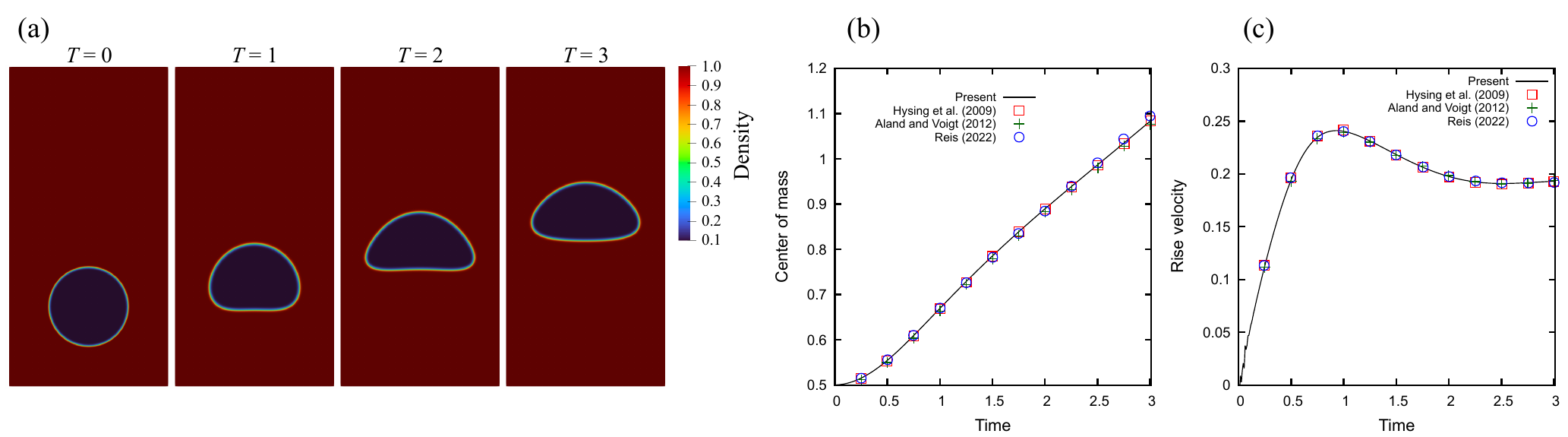}
    \caption{Bubble rising simulation for Case 1 ($Re=35$, $Eo=10$, $\rho_r^0/\rho_b^0=10$, and $\mu_r/\mu_b=10$) in Table~\ref{tab:2dBubbleRisingParameters}. 
    Results from finite-element-method simulations with level-set-based approach (with FreeLIFE solver)~\citep{Hysing2009-cd} and phase-field-based approach~\citep{Aland2012-ip} and the phase-field LB model~\citep{Reis2022-lo} are also shown.
    }
    \label{fig:2D_bubble(Case1)}
\end{figure*}

Figure~\ref{fig:2D_bubble(Case1)} shows the time evolution of the bubble shape, center of mass, and rise velocity obtained from the simulation results for Case 1.
In this case, the domain was discretized into $N_x \times N_y = 160 \times 320$, and we set $\beta = 0.7$.
Using the parameter set in Table~\ref{tab:2dBubbleRisingParameters}, the Mach number with respect to the speed of sound in the liquid phase was obtained as $Ma = U/c_s^r = 0.00548$ and the number of iterations up to $T=3$ was calculated to be $364\,000$.
As shown in Fig.~\ref{fig:2D_bubble(Case1)}(a), the initially circular bubble rises under the effect of gravity and is deformed by the balance between gravity and surface tension.
Similar to the benchmark solution~\citep{Hysing2009-cd}, no bubble breakup occurs.
In Fig.~\ref{fig:2D_bubble(Case1)}(b) and (c), the present results are compared with available numerical data: finite-element-method simulations with a level-set-based approach (with FreeLIFE solver)~\citep{Hysing2009-cd} and phase-field-based approach~\citep{Aland2012-ip} and the phase-field LB model~\citep{Reis2022-lo}.
% The bubble shape at $T=3$ is qualitatively in good agreement with all existing studies. 
The time histories of the center of mass and rise velocity are in good agreement with those of all existing studies.

\begin{figure*}
    \centering
    \includegraphics[width=1\linewidth]{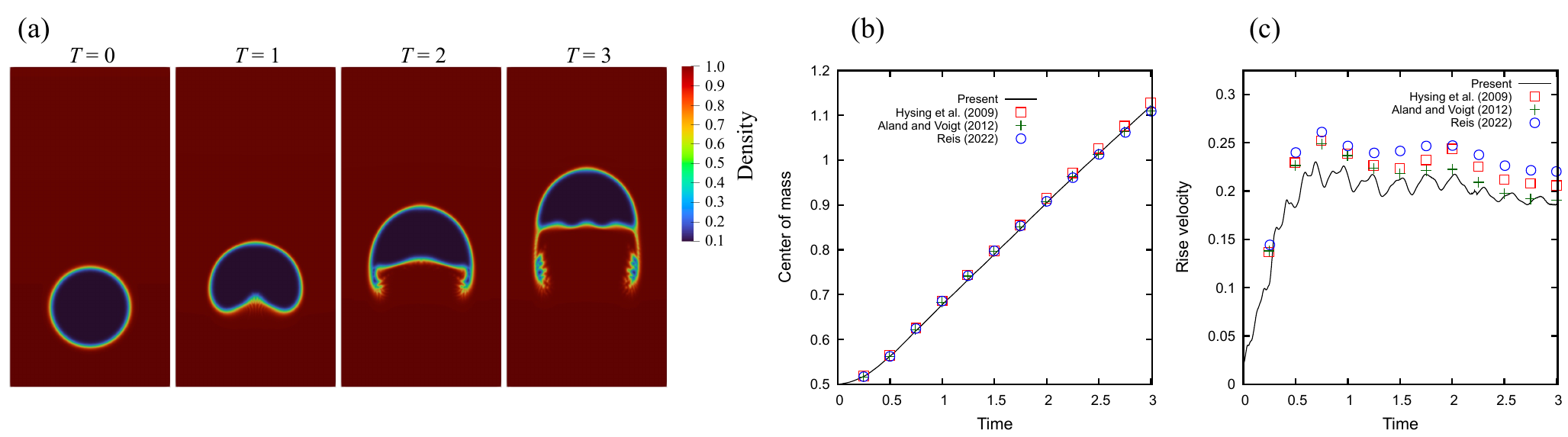}
    \caption{Bubble rising simulation for Case 2 ($Re=35$, $Eo=125$, $\rho_r^0/\rho_b^0=1000$, and $\mu_r/\mu_b=100$) in Table~\ref{tab:2dBubbleRisingParameters}.  
    Results from finite-element-method simulations with level-set-based approach (with FreeLIFE solver)~\citep{Hysing2009-cd} and phase-field-based approach~\citep{Aland2012-ip} and the phase-field LB model~\citep{Reis2022-lo} are also shown.
    }
    \label{fig:2D_bubble(Case2)}
\end{figure*}

Figure~\ref{fig:2D_bubble(Case2)} presents the simulation results for Case 2.
In this case, the domain was discretized into $N_x \times N_y = 320 \times 640$.
After several test simulations, we set $\beta = 0.2$ to widen the range of interfaces.
Using the parameter set in Table~\ref{tab:2dBubbleRisingParameters}, the Mach number with respect to the speed of sound in the liquid phase was obtained as $Ma = U/c_s^r = 0.0548$ and the number of iterations up to $T=3$ was calculated to be $728\,000$.
From Fig.~\ref{fig:2D_bubble(Case2)}(a), it can be verified that the simulation is stable and does not break at a density ratio of $1\,000$. 
At $T=3$, the ends of the bubble appear to break, similar to that observed in previous studies ~\citep{Hysing2009-cd,Reis2022-lo}.
From Fig.~\ref{fig:2D_bubble(Case2)}(b) and (c), it can be concluded that the time history of the center of mass agrees well with that observed in the existing studies; however, the discrepancy is large with respect to the rising velocity.
There are two main reasons for this finding.
The first corresponds to the definition of the bubble velocity given in Eq.~(\ref{eq:bubble_rise_velocity}). 
To manage high-density ratios, the interface was thickened by setting $\beta = 0.2$, and the velocity in the interfacial region was included in the evaluation within the definition of Eq.~(\ref{eq:bubble_rise_velocity}), which decreased the overall rise velocity.
The second factor is the influence of the compressibility of the liquid phase.
In a typical CG model, the speed of sound in the liquid phase decreases with an increasing density ratio because of the assumption of an ideal gas EOS for each phase.
A reduction in the speed of sound in the liquid phase at such high-density ratios can be prevented by introducing a suitable EOS~\citep{Lafarge2021-ce} to replace the ideal gas EOS.
Nevertheless, to the best of our knowledge, this presents a novel approach to solving Case 2 using the CG LB model.

% Fig.~\ref{fig:setup_2d_bubble}.
% \begin{figure}[tb]
%     \centering
%     \includegraphics[width=0.8\linewidth]{figures/setup_2d_bubble.pdf}
%     \caption{Setup for bubble rising.}
%     \label{fig:setup_2d_bubble}
% \end{figure}


\subsection{Three-dimensional Rayleigh--Taylor instability}

The final numerical experiment involves the three-dimensional Rayleigh--Taylor instability.
The Rayleigh--Taylor instability is a fundamental interfacial instability induced when a heavy fluid is placed over a light fluid subjected to a slightly disturbed interface in a gravitational field~\citep{Chandrasekhar1961-ly}.
Under certain conditions, Kelvin--Helmholtz instability occurs owing to the velocity difference across the fluid--fluid interface.
This problem involves complex interface deformation.

We refer to the computational setup adopted in a previous study \citet{He1999-pp}.
A schematic of the computational setup is shown in Fig.~8 of Ref.~\citep{Saito2017-lg}.
The top and bottom boundaries are no-slip walls and the lateral boundaries are periodic.
As described in Ref.~\citep{He1999-pp}, a single-mode initial perturbation is imposed as:
\begin{equation}
    h(x,y) = 0.05 W \qty[ \cos (\frac{2\pi x}{W}) + \cos (\frac{2\pi y}{W}) ],
\end{equation}
in the mid-plane, where $W$ is the computational domain width.
The body force for this problem is incorporated as:
\begin{equation}
    \vb{F}(\vb{x},t) = -\qty[ \rho(\vb{x},t) - \frac{\rho_r^0 + \rho_b^0}{2} ] \vb{g},
\end{equation}
where $\vb{g}=(0,0,-g)$.
The gravitational acceleration $g$ is chosen to satisfy the relation $(Wg)^{1/2}=0.04$~\citep{He1999-pp}.
The computational domain is set as $W\times W \times 4W$ with $W=128$.
The Atwood number,
\begin{equation}
    At = \frac{\rho_r^0 - \rho_b^0}{\rho_r^0 + \rho_b^0},
\end{equation}
which is a dimensionless number that characterizes this problem, is fixed at 0.5 throughout the simulations. 
This corresponds to a density ratio of 3.
Interface tension is neglected; thus, the perturbed interface is expected to always be unstable in the inviscid case.
The kinematic-viscosity ratio is set to unity.
Another dimensionless number is the Reynolds number, which is defined as:
\begin{equation}
    Re = \frac{\sqrt{Wg}W}{\nu}.
\end{equation}
In this problem, time is scaled by $(W/g)^{1/2}$.

\begin{figure}[tb]
    \centering
    \includegraphics[width=1\linewidth]{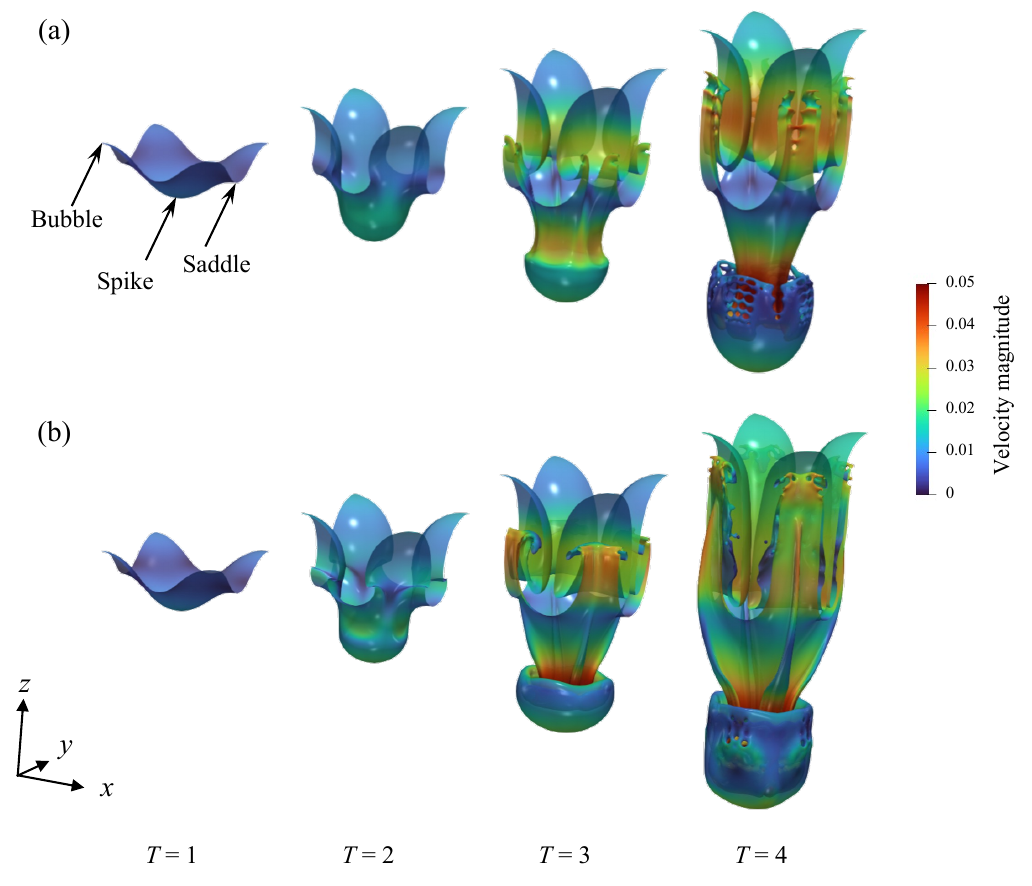}
    \caption{Interface evolution of three-dimensional Rayleigh--Taylor instability for $Re=1024$ and $At=0.5$.
    The original equilibria for the CG model are used for the top panels (Model A in Table~\ref{tab:summari_of_equilibria}), while the generalized equilibria proposed in this study are used for the bottom panels (Model D in Table~\ref{tab:summari_of_equilibria}).
    }
    \label{fig:RTI_comparison_interface}
\end{figure}

Figure~\ref{fig:RTI_comparison_interface} shows the time evolution of the interface, that is, the isosurface with $\phi=0$, obtained from the simulations for $Re=1\,024$. 
The upper panels [Fig.~\ref{fig:RTI_comparison_interface}(a)] show the results obtained using the original equilibria (Model A in Table~\ref{tab:summari_of_equilibria}),
whereas the lower panels [Fig.~\ref{fig:RTI_comparison_interface}(b)] show the results obtained using the generalized equilibria (Model D in Table~\ref{tab:summari_of_equilibria}).
Surfaces regarded as interfaces are colored according to the velocity magnitude.
In both cases, the overall trend of the interface shape is similar. For example, the spike tip descends over time to form a mushroom-like roll-up.
In the initial stage, at $T=1$, no significant difference is observed in the change in the shape of the interface because of the equilibrium employed.
Focusing on the later stages, after $T=2$, differences are observed in both the interface shape and velocity, depending on the equilibria used.
This is because of the difference in accuracy owing to the applied equilibria, as discussed in Secs.~\ref{sec:moving_droplet} and \ref{sec:layered-two-phase}; the generalized equilibria proposed in this study accurately capture the velocity distribution near the interface, whereas the original equilibria result in an unphysical velocity discontinuity near the interface.
Over time, these differences in behavior have a greater effect on the interface shape and velocity field.

\begin{figure}[tb]
    \centering
    \includegraphics[width=0.8\linewidth]{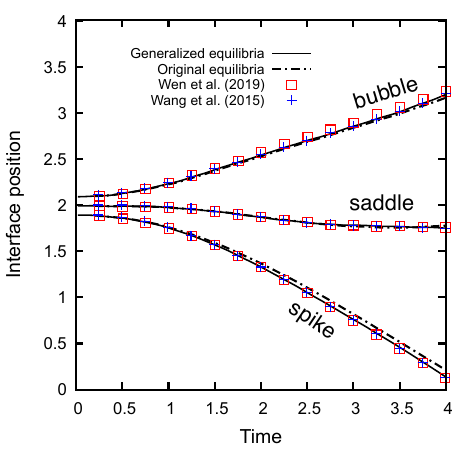}
    \caption{Time histories of interface position for $At=0.5$ and $Re=1024$ with generalized (solid line) and original equilibria (broken line). 
    Results from the improved CG RM-MRT model~\citep{Wen2019-jc} and multiphase LB flux solver~\citep{Wang2015-an} are also shown.}
    \label{fig:RTI_TimeHistory}
\end{figure}

To better quantify the time evolution of the Rayleigh--Taylor instability, the interfacial position was measured, and the results are shown in Fig.~\ref{fig:RTI_TimeHistory}.
Measurements were taken at three characteristic points: the bubble, saddle, and spike (their locations are shown in Fig.~\ref{fig:RTI_comparison_interface}).
The result of the generalized equilibria is shown as a solid line, whereas the result of the original equilibria is shown as a broken line.
The difference between the generalized equilibria (solid line) and original equilibria (broken line) is particularly noticeable in the time history of the spike.
The original equilibria exhibit a slower evolution of the tip position.
The velocity discontinuity near the interface is considered to have prevented the development of the spike tip.
In addition, the data obtained by the improved CG LB~\citep{Wen2019-jc} and multiphase LB flux solver~\citep{Wang2015-an} are shown in Fig.~\ref{fig:RTI_TimeHistory} for comparison with previous numerical simulations.
In regards to the saddle point and spike tip, the present generalized equilibria results are in good agreement with those in Refs.~\citep{Wang2015-an,Wen2019-jc}, with no noticeable differences.
In regards to the bubble tip, the generalized equilibrium LB and multiphase LB flux solver results are in good agreement, although the improved CG model progresses slightly faster.

% \begin{figure}
%     \centering
%     \includegraphics[width=0.7\linewidth]{figures/RTI_comparison_interface_velocity.pdf}
%     \caption{at $T=4$.}
%     \label{fig:RTI_comparison_interface_velocity}
% \end{figure}

\begin{figure}[htbp]
    \centering
    \includegraphics[width=1\linewidth]{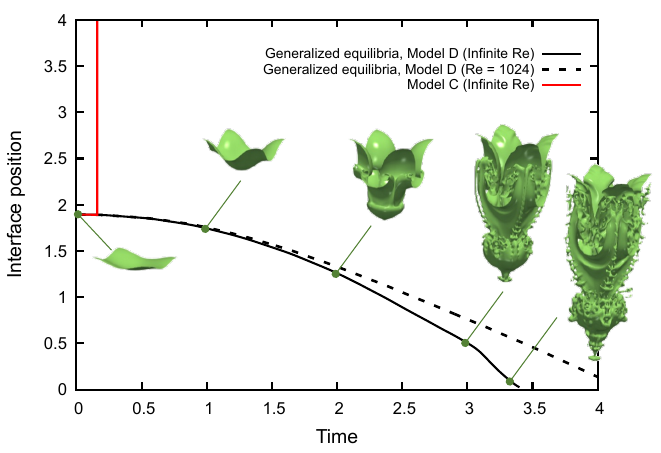}
    \caption{Time history of spike position during Rayleigh--Taylor instability with images of interfaces at infinite Reynolds number with zero viscosity.}
    \label{fig:RTI_Re_Inf}
\end{figure}

The following discussion focuses on whether any difference can be observed between the simulations of Models C and D (Table~\ref{tab:summari_of_equilibria}).
Based on the numerical experiments presented in Secs. ~\ref{sec:moving_droplet} and \ref{sec:layered-two-phase}, it can be concluded that their computational accuracies are nearly equivalent.
For more details, additional simulations of the Rayleigh-Taylor instability were performed.
The same settings as those in Fig.~\ref{fig:RTI_comparison_interface} were used, and only the kinematic viscosity was changed to $\nu=0$.
The simulations were performed at an infinite Reynolds number.
Figure~\ref{fig:RTI_Re_Inf} shows the time history of the spike tip at an infinite Reynolds number.
The results for $Re = 1\,024$ are also presented for reference.
Images of the interface shape at infinite Reynolds numbers are shown.
We observe that the evolution of the spike tip is considerably larger as the Reynolds number is increased. Furthermore, the computation is stable.
From the interface shape, it can be seen that the Kelvin--Helmholtz instability at the interface is particularly prominent in the later stages.
However, in the case of Model C (red solid line), the computation was broken, even though it was computed within the same CM-MRT framework.
This indicates that the velocity-dependent components of the equilibrium CMs in Eqs.~(\ref{eq:equilibrium_central_moments_improved3rd_4th})--(\ref{eq:equilibrium_central_moments_improved3rd_6th}) contribute to the numerical instability of the computation. The proposed generalized equilibrium CMs [Eqs.~(\ref{eq:equilibrium_central_moments_proposed_0th})--(\ref{eq:equilibrium_central_moments_proposed_6th})] is effective in avoiding this numerical instability.
It should be noted that the ability to stably simulate does not necessarily imply an accurate solution, as discussed in \citet{De_Rosis2017-qm}.
In simulations involving high Reynolds numbers, choosing appropriate values for both the mesh size and kinematic fluid viscosity is of paramount importance.
In conclusion, the equilibria proposed in this study not only provide simplified velocity-independent forms in the CM space but also benefit from the high numerical stability of the CM-MRT and can stably solve multiphase flows with extremely high Reynolds numbers.

% \textcolor{red}{
% Accurate numerical calculations of high Reynolds flow fields can only be performed if the adopted grid resolution is fine enough to predict the rise of small-scale structures typical of turbulent domains [49]. 
% Moreover, as the kinematic viscosity of the fluid approaches the vanishing limit, the role of the numerical viscosity becomes increasingly prominent and dominant, leading to unrealistic results [32], where the numerical viscosity inherent in the LB scheme eventually takes over the kinematic fluid viscosity when the latter vanishes. In short, one should be particularly careful when simulating a high Reynolds scenario, taking care to adopt appropriate values of both the grid size and the kinematic fluid viscosity.
% }

\section{Conclusions\label{sec:conclusion}}

In this study, we propose the generalized equilibria for a three-dimensional CG LB model and investigate its numerical properties within the framework of the CM-MRT model.

First, the equilibrium distribution function of the CG model is reformulated in a more general form using Hermite polynomials [Eq.~(\ref{eq:equilibria_in_general_form})], which can provide a prospective indication of the deviation from the ideal gas scenario.
An examination of the equilibrium distribution function and its CMs in the existing CG model shows that the velocity dependence of the equilibrium CM vanishes in the order of the single-phase equilibrium distribution function $g_i^{\mathrm{eq},N}$ and the correction operator $\Phi_i$.
Inspired by this fact, we formulate an equilibrium distribution function in Hermite polynomials up to the sixth order for both $g_i^{\mathrm{eq},N}$ and $\Phi_i$.
In the phase space, this equilibrium distribution function contains terms up to $O(u^6)$, increasing its complexity compared to its original form, and its implementation becomes more cumbersome.
However, most equilibrium CMs are zero, and the nonzero moments have simple functional forms with no velocity dependence.

Numerical experiments show that our approach improves the Galilean invariance, and the accuracy for static problems is comparable to that of the previous approach~\citep{Ba2016-ve,Wen2019-jc}, considering up to the third order for $\Phi_i$.
In a dynamic problem, the bubble rise benchmark set by \citet{Hysing2009-cd} was applied and was found to be as accurate as that of other methods for a density ratio of 10.
In contrast, at a density ratio of $1\,000$, the center of mass was comparable to that obtained with other studies, but the rise velocity was different.
The numerical stability of this dynamic and high-density-ratio problem is noteworthy; however, further improvements are required to solve it more accurately within the framework of the CG model.

Finally, simulations of the three-dimensional Rayleigh--Taylor instability were performed under the condition of $At=0.5$.
In the simulations for $Re = 1\,024$, a large difference is observed in the time evolution of the interface shape and tip position with and without the Galilean invariance correction.
Furthermore, simulations with zero kinematic viscosity (infinite Reynolds number) show that the equilibrium CMs above the fourth order significantly contribute to the numerical stability.

The concept of the proposed generalized equilibria is applicable to the forcing-based free-energy LB model~\citep{Li2021-ar}.
Because the equilibrium CMs have simplified forms, those in the cumulant space are similarly simple. 
Therefore, the present CM-MRT-based model can be easily extended to a cumulant-based model~\cite{Geier2015-pv}.
Furthermore, by following the strategy of \citet{De_Rosis2020-uj} starting from equilibrium distribution functions proposed in Ref.~\citep{Coreixas2019-ce}, the present framework could be implemented on more compact lattices (e.g., D3Q19) to improve the computational efficiency.

\begin{acknowledgments}
This work was supported by JSPS KAKENHI Grant Numbers JP20K04297 and JP22K14201.
Part of this work was supported by the New Energy and Industrial Technology Development Organization (NEDO) JPNP14004.
\end{acknowledgments}

\appendix

\section{Generalized equilibria for D2Q9 lattice}
For a typical D2Q9 lattice in two dimensions, the lattice velocity $\vb{c}_i$ is defined as~\citep{He1997-ur}:
\begin{equation}
\mathbf{c}_i 
=\begin{bmatrix}
   c_{ix}    \\
   c_{iy}    
\end{bmatrix}
= c
\begin{bmatrix}
   0 & 1 & -1 & 0 & 0 & 1 & -1 & 1 & -1   \\
   0 & 0 & 0 & 1 & -1 & 1 & -1 & -1 & 1   
\end{bmatrix},
\end{equation}
The weight function is expressed as follows:
\begin{equation}
    w_i = 
    \begin{cases}
        4/9, & |\vb{c}_i| = 0 \\
        1/9, & |\vb{c}_i| = 1 \\
        1/36. & |\vb{c}_i| = \sqrt{2} 
    \end{cases}
\end{equation}
The lattice speed of sound is the same as D3Q27: $c_s^2=1/3$.
Using Hermite polynomials up to the fourth order, the equilibrium distribution function for single-phase can be expressed as follows ~\citep{De_Rosis2019-bi}:
\begin{equation}
  \begin{split}
    g_i^{\mathrm{eq,4}} = & ~ \rho w_i \left(  1 + 
    %% 1st order
      \frac{u_x H_{i10} + u_y H_{i01}}{c_s^2} \right.  \\
    %% 2nd order 
    & \left. 
    + \frac{u_x^2 H_{i20} + u_y^2 H_{i02}
      + 2u_x u_y H_{i11}}{2c_s^4} \right.\\
    %% 3rd order
    & \left.  + \frac{
        u_x^2 u_y H_{i21} 
      + u_x u_y^2 H_{i12}}{2c_s^6} 
    %% 4th order
      + \frac{u_x^2 u_y^2 H_{i22}
        }{4c_s^8} \right), 
  \end{split}
\end{equation}
with Hermite polynomials:

\noindent 
Zeroth order:
\begin{equation}
    H_{i00} = 1,
\end{equation}
First order:
\begin{equation}
    H_{i10} = c_{ix},~
    H_{i01} = c_{iy},~
\end{equation}
Second order:
\begin{equation}
\begin{split}
    H_{i20} = &~ c_{ix}^2 - c_s^2, \\
    H_{i02} = &~ c_{iy}^2 - c_s^2, \\
    H_{i11} = &~ H_{i10} H_{i01},\\
\end{split}
\end{equation}
Third order:
\begin{equation}
\begin{split}
    H_{i12} = &~ H_{i10} H_{i02}, \\
    H_{i21} = &~ H_{i20} H_{i01}, \\
\end{split}
\end{equation}
Fourth order:
\begin{equation}
\begin{split}
    H_{i22} = &~ H_{i20} H_{i02}. \\
\end{split}
\end{equation}

The isotropic operator for the D2Q9 lattice is expressed as:  
\begin{equation}
    E_i = w_i 
    \qty(\frac{H_{i20}+H_{i02}}{2c_s^4} 
    - \frac{H_{i22}}{4c_s^6}
    ),
\end{equation}
which is equivalent to Eq.~(40) in Ref.~\citep{Lafarge2021-ce}.

The generalized equilibrium distribution function for the D2Q9 lattice is expressed as:
\begin{widetext}
\begin{equation}
  f_i^{\mathrm{eq}} = g_i^{\mathrm{eq},4} + 
    ( p - \rho c_s^2 ) 
    \left( E_i  + w_i \left[ \frac{u_x H_{i12} + u_y H_{i21}}{2c_s^6} + \frac{(u_x^2 + u_y^2)  H_{i22} )}{4c_s^8}   
     \right] \right), 
     \label{eq:equilibria_in_general_form_in_2D}
\end{equation}
\end{widetext}
The equilibrium CMs are computed as follows: 
\begin{equation}
    k_{\alpha\beta}^{\mathrm{eq}} = \sum_i {
    f_i^{\mathrm{eq}} 
    (c_{ix} - u_x)^\alpha
    (c_{iy} - u_y)^\beta}.
    \label{eq:definition_equilibrium_central_moments_in_2d}
\end{equation}
Consequently, the equilibrium CMs in Eq.~(\ref{eq:equilibria_in_general_form_in_2D}) are obtained as:

\noindent Zeroth order: 
\begin{equation}
    k_{00}^{\mathrm{eq}} =  \rho,
\end{equation}
First order:
\begin{equation}
  k_{10}^{\mathrm{eq}} = k_{01}^{\mathrm{eq}} = 0  , 
\end{equation}
Second order:
\begin{equation}
  \begin{split}
  k_{11}^{\mathrm{eq}}  &= 0  ,  \\
  k_{20}^{\mathrm{eq}} = k_{02}^{\mathrm{eq}} &= p  ,  \\
  \end{split}
\end{equation}
Third order:
\begin{equation}
  \begin{split}
  % 3rd order
  k_{12}^{\mathrm{eq}} = k_{21}^{\mathrm{eq}}  = k_{11}^{\mathrm{eq}} = 0, 
  \end{split}
\end{equation}
Fourth order:
\begin{equation}
  % 4th order
  k_{22}^{\mathrm{eq}} =  p c_s^2.  \\
\end{equation}

\section{Hermite polynomials in three dimensions\label{sec:Appendix_Hermite_polynomials}}
Following \citet{De_Rosis2019-bi}, the Hermite polynomials for three dimensions are expressed as follows:

\noindent 
Zeroth order:
\begin{equation}
    H_{i000} = 1,
\end{equation}
First order:
\begin{equation}
    H_{i100} = c_{ix},~
    H_{i010} = c_{iy},~
    H_{i001} = c_{iz},
\end{equation}
Second order:
\begin{equation}
\begin{split}
    H_{i200} &= c_{ix}^2 - c_s^2, 
    ~H_{i020} = c_{iy}^2 - c_s^2, 
    ~H_{i002} = c_{iz}^2 - c_s^2, \\
    H_{i110} &= H_{i100} H_{i010},
    ~H_{i011} = H_{i010} H_{i001},
    ~H_{i101} = H_{i100} H_{i001},
\end{split}
\end{equation}
Third order:
\begin{equation}
\begin{split}
    H_{i120} &= H_{i100} H_{i020}, 
    ~H_{i102} = H_{i100} H_{i002}, \\
    H_{i012} &= H_{i010} H_{i002}, 
    ~H_{i210} = H_{i200} H_{i010}, \\
    H_{i201} &= H_{i200} H_{i001}, 
    ~H_{i021} = H_{i020} H_{i001}, \\
    H_{i111} &= H_{i100} H_{i010}H_{i001}, \\
\end{split}
\end{equation}
Fourth order:
\begin{equation}
\begin{split}
    H_{i220} = &~ H_{i200} H_{i020}, \\
    H_{i202} = &~ H_{i200} H_{i002}, \\
    H_{i022} = &~ H_{i020} H_{i002}, \\
    H_{i211} = &~ H_{i200} H_{i010} H_{i001}, \\
    H_{i121} = &~ H_{i100} H_{i020} H_{i001}, \\
    H_{i112} = &~ H_{i100} H_{i010} H_{i002}, 
\end{split}
\end{equation}
Fifth order:
\begin{equation}
\begin{split}
    H_{i122} = &~ H_{i100} H_{i020} H_{i002}, \\
    H_{i212} = &~ H_{i200} H_{i010} H_{i002}, \\
    H_{i221} = &~ H_{i200} H_{i020} H_{i001},
\end{split}
\end{equation}
Sixth order:
\begin{equation}
    H_{i222} = H_{i200}H_{i020}H_{i002}.
\end{equation}

% \nocite{*}

\bibliography{aps}% Produces the bibliography via BibTeX.

\end{document}